\documentclass[nofootinbib,preprintnumbers,showpacs]{revtex4}
\usepackage{graphicx}
\usepackage{dcolumn}
\usepackage{amsmath,amssymb,epsfig}
\usepackage{paralist}
\usepackage{comment}
\usepackage{graphicx}
\usepackage{multirow}
\usepackage{wrapfig}
\usepackage{color,soul}
\usepackage[normalem]{ulem}

\makeatletter
\def\squiggly{\bgroup \markoverwith{\textcolor{red}{\lower3.5\p@\hbox{\sixly \char58}}}\ULon}
\makeatother

\renewcommand{\vec}[1]{\boldsymbol{\mathrm{#1}}}

\newcommand{\editnote}[2]{}

\allowdisplaybreaks

\begin{document}

\title{General relativistic observables for the ACES experiment}

\author{Slava G. Turyshev$^1$, Nan Yu$^1$, and Viktor T. Toth$^2$}

\affiliation{\vskip 3pt
$^1$
Jet Propulsion Laboratory, California Institute of Technology, 4800 Oak Grove Drive, Pasadena, CA 91109-0899, USA
}%

\affiliation{\vskip 3pt
$^2$Ottawa, ON  K1N 9H5, Canada
}%

\date{\today}

\begin{abstract}

We develop a high-precision model for relativistic observables of the Atomic Clock Ensemble in Space (ACES) experiment on the International Space Station (ISS). We develop all relativistic coordinate transformations that are needed to describe the motion of ACES in Earth orbit and to compute observable quantities. We analyze the accuracy of the required model as it applies to the proper-to-coordinate time transformations, light time equation, and spacecraft equations of motion. We consider various sources of nongravitational noise and their effects on ACES. We estimate the accuracy of orbit reconstruction that is needed to satisfy the ACES science objectives. Based on our analysis, we derive models for the relativistic observables of ACES, which also account for the contribution of atmospheric drag on the clock rate. We include the Earth's oblateness coefficient $J_2$ and the effects of major nongravitational forces on the orbit of the ISS. We demonstrate that the ACES reference frame is pseudo-inertial at the level of accuracy required by the experiment. We construct a Doppler-canceled science observable representing the gravitational redshift. We derive accuracy requirements for ISS navigation. The improved model is accurate up to $<1$~ps and $\sim 4\times 10^{-17}$ for time and frequency transfers, correspondingly. These limits are determined by the higher order harmonics in Earth's gravitational potential.

\end{abstract}


\maketitle


\section{Introduction}
\label{sec:intro}

The Atomic Clock Ensemble in Space (ACES) experiment \cite{Cacciapuoti-Salomon:2011} is developed by the ESA\footnote{The European Space Agency, {\tt http://www.esa.eu/}} and CNES\footnote{The Centre National d'Etudes Spatiales (CNES) -- the French Space Agency, {\tt http://www.cnes.fr/}}, to be flown on the International Space Station (ISS) in 2017--18. The ACES experimental package consists of two atomic clocks: the cold-atom clock PHARAO\footnote{Projet d'Horloge Atomique Par Refroidissement d'Atomes En Orbite (PHARAO), https://pharao.cnes.fr/en/PHARAO/index.htm} (developed by CNES \cite{Laurent-etal:2008}) and a space hydrogen maser (SHM, developed by SpectraTime SA\footnote{SpectraTime SA, see {\tt http://www.spectratime.com/}} \cite{Goujon-etal:2010}). The experiment will be attached to the external payload facility on the European Columbus module on the space station. Associated with ACES is the European Laser Timing (ELT) experiment \cite{Prochazka-etal:2013}.

Placed in a microgravity environment on the ISS, the expected overall frequency stability of PHARAO is $1\times 10^{-16}$ (see Table~{\ref{tb:caps}). The short-term frequency stability of PHARAO will be evaluated by direct comparison to the SHM. Long term stability will be measured by comparison to ultrastable ground clocks and systematic frequency shifts will be evaluated {\em in situ}. The medium term frequency instability will be evaluated by direct comparison to ultra-stable ground clocks. The long-term stability will be determined by on-board comparison to PHARAO.

To compare time and frequency between various ground clocks, ACES will use a two-way microwave system called the microwave link (MWL) \cite{Schaefer-etal:2008,Schreiber-etal:2009,Delva-etal:2012} (Table~\ref{tb:caps}). Given the anticipated accuracy of the MWL, ACES is expected to provide absolute synchronization of ground clock time scales with an uncertainty of 100~ps. The experiment will also enable  comparison of primary frequency standards with accuracy at the $10^{-16}$ level. Additionally, high precision time transfer will be facilitated by the ELT experiment, with overall planned accuracy of 50~ps, and a per-pass space-to-ground clock comparison precision of 4~ps.

With an atomic clock offering such accurate performance in the microgravity environment, ACES will conduct several tests of fundamental physics. Specifically, ACES will conduct gravitational redshift measurements, test Lorentz-invariance, and it will search for possible variations in the fine structure constant. It is expected that the uncertainty on the gravitational redshift measurement will be below $50 \times 10^{-6}$ for an integration time corresponding to one ISS pass. However, with the ultimate accuracy of PHARAO, ACES may reach an uncertainty level of $2\times 10^{-6}$. Measurements can reach a precision level of $\delta c / c \simeq10^{-10}$ in the search for anisotropies of the speed of light. Time variations of the fine structure constant, $\alpha$, can be measured at the level of precision $\dot\alpha/\alpha < 1 \times 10^{-16}$ yr$^{-1}$.

\begin{table}[t]
\caption{\label{tb:caps} Anticipated frequency and timing stability of the ACES package \cite{Cacciapuoti-Salomon:2011,Duchayne-etal:2007,Duchayne-etal:2009,ESA-ACES}. Note that one pass is $\sim$300 sec.}
\begin{tabular}{|l|c|c|c|c|c|c|}\hline
ACES performance parameter&1~s&300~s&$10^3$~s&$10^4$~s&1~day&10~days\\
\hline\hline
PHARAO frequency stability&$10^{-13}$&~&~&~&$3\times 10^{-16}$&$10^{-16}$\\
SHM frequency stability&~&~&$2.1\times 10^{-15}$&$1.5\times 10^{-15}$&~&~\\
Common view comparison&~&~&2~ps&5~ps&20~ps&~\\
Non-common view comparison&~&0.3~ps&~&~&6~ps&23~ps\\\hline
\end{tabular}
\end{table}

To reach its science objectives, ACES will rely on the time and frequency stability of the SHM, PHARAO, the MWL and clocks over one ISS pass. The most important component of a clock is an oscillator, the periodic oscillation of which has to be generated, maintained and read out by suitable means. As it is known, an oscillator on board the ISS is subject to many classical disturbances. In addition to relativistic gravity, clock performance is affected by nongravitational forces external to the station and also by the dynamical environment on the ISS. This environment is rather complex and includes forces and torques due to causes such as extended structure vibrations, frequent thruster firings, spacecraft docking and undocking, on-going human activity, thermal imbalance, outgassing, etc. Some of these effects directly impact the trajectory of the ISS and consequently, the relativistic timing and frequency transfer observables of ACES. Other effects will directly impact the clock stability by producing unwanted acceleration noise at the clock's location. These effects should be accounted for in a classical description of an oscillator subjected to various sources of acceleration noise.

Although a Newtonian formulation of the coordinate reference systems for the ISS is readily available \cite{SSP-30219J}, it is not sufficient for ACES. The requirement to formulate models of ACES observables within the framework of Einstein's general theory of relativity was recognized early on during mission development \cite{Blanchet-etal:2001,Petit-Wolf:2005}. ACES will rely on accurate navigation of the ISS and accurate timing measurements between the station and ground-based terminals.  The experiment will require precision timing of all critical events related to the transmission and reception of various microwave and optical signals used on ACES for time and frequency transfer, as well as navigation. The resulting time series of high accuracy radio-metric and opto-metric data will provide the time transfer accuracy that is needed to perform tests of fundamental physics with ACES. Based on the anticipated performance of hardware that will be involved in the ACES experiment, the models for the ACES observables must be accurate at the level of 1~ps and $\delta f/f\simeq 1\times 10^{-16}$ for time and frequency transfers, correspondingly. We will use these numbers in developing relativistic models for observables on the ACES experiment.

To describe the dynamics around the Earth we will introduce several reference frames, each with its own coordinate chart.
In the immediate vicinity of the Earth we can introduce a set of {\it local} coordinates defined in the frame associated with the Earth: The origin of the Geocentric Coordinate Reference System (GCRS) is the Earth's center of mass. Positions of ground stations are given with respect to another terrestrial coordinate system, the Topocentric Coordinate Reference System (TCRS; see also Ref.~\cite{IERS2010}). We also consider the Spacecraft Coordinate Reference System (SCRS), the origin of which is fixed at the ISS' center of mass. We also use ACES Coordinate Reference System (ACRS) associated with the ACES package. The definition and properties of the TCRS, together with useful details on relativistic time-keeping in the solar system are given in \cite{Turyshev-etal:2012}. The SCRS was discussed in \cite{Turyshev-etal:2012} in the context of the GRAIL mission. Here we introduce the SCRS on the ISS and we define the ACRS together with appropriate coordinate transformations.

Previous studies of ACES \cite{Blanchet-etal:2001,Petit-Wolf:2005,Duchayne-etal:2009,Wermuth-etal:2012} offered relativistic models of the basic experimental observables: time and frequency transfer. Some of these efforts treated the ISS as a free-falling platform, moving on a geodesic worldline in the GCRS. However, the motion of the space station is subject to nongravitational forces that are present in the near-Earth environment (as discussed in \cite{Duchayne-etal:2009,Ashby-etal-2014}), including atmospheric drag, solar radiation pressure and thermo-elastic cycling. As the shape of the space station is complex, its center of mass does not coincide with its center with respect to nongravitational forces. This mismatch results in torques that affect the orientation of the ISS. Also, the presence of these dissipative forces results in the ISS constantly loosing altitude, in apparent violation of naive models of energy-momentum conservation. Furthermore, as the ISS orbits the Earth, its extended structure vibrates and flexes, resulting in a complex profile of nongravitational acceleration noise on the station.

This paper is organized as follows: In Section \ref{sec:frames} we discuss the conventional definitions of the GCRS and TCRS, including representations of the metric tensor in these reference systems and the coordinate transformations between them. We also present the equations of motion for Earth-orbiting spacecraft and light-time equations.  We pay special attention to contributions by various nongravitational forces acting on the ISS and the ACES package, including atmospheric drag and solar radiation pressure. In Section~\ref{sec:crs-ISS} we present the coordinate reference frames important for the ACES experiment, namely the SCRS and the ACRS, again including representations of the metric tensor and coordinate transformations, at a level of accuracy appropriate for ACES. In Section~\ref{sec:obsere} we discuss the formulation of the relativistic observables of the ACES experiment. We present models for time and frequency transfer and evaluate the navigational requirements needed to fulfil the science objectives of ACES. We conclude with a set of recommendations and an outlook in Sec.~\ref{sec:conc}.

\begin{figure}
\includegraphics[width=0.45\linewidth]{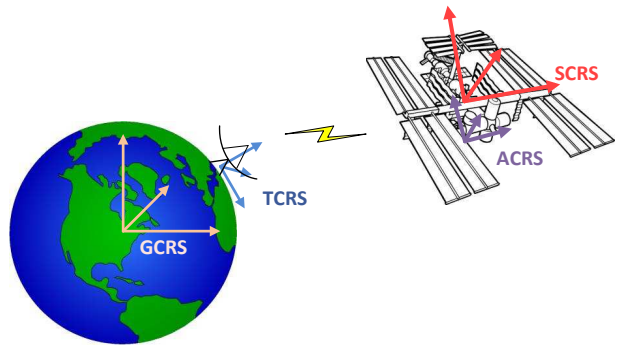}
\caption{Schematic relationship of the coordinate systems discussed in the text(not to scale).
}
\end{figure}

\section{Reference frames and the equations of motion}
\label{sec:frames}

A properly defined set of coordinate reference frames simplifies the discussion of the observables of an experiment. Because of its high-precision science objectives and its deployment on the ISS, ACES presents a set of unique features. One of the primary challenges is the presence of various nongravitational forces acting on the experimental payload. In this section we present descriptions of various reference frames relevant to ACES. We pay particular attention to the dynamical environment relevant to each of these reference frames. Although our analysis is motivated by the ACES experiment, the results obtained here are relevant to other missions that will use precision clocks to reach their science objectives \cite{SOC}.

The coordinate systems that represent the ACES experiment are each characterized by a unique set of harmonic potentials including the scalar potential $w$ and the vector potential $w^\lambda$. Using these harmonic potentials, we represent the metric tensor of a reference frame in the form\footnote{
The notational conventions used in this paper are as follows. Latin indices ($m,n,...$) are space-time indices that run from 0 to 3. Greek indices $\alpha,\beta,...$ are spatial indices that run from 1 to 3. In case of repeated indices in products, the Einstein summation rule applies: e.g., $a_mb^m=\sum_{m=0}^3a_mb^m$. Bold letters denote spatial (three-dimensional) vectors: e.g., ${\vec a} = (a_1, a_2, a_3), {\vec b} = (b_1, b_2, b_3)$. The dot is used to indicate the Euclidean inner product of spatial vectors: e.g., $(\vec a \cdot \vec b) = a_1b_1 + a_2b_2 + a_3b_3$. Latin indices are raised and lowered using the metric $g_{mn}$. The Minkowski (flat) space-time metric is given by $\gamma_{mn} = {\rm diag} (1, -1, -1, -1)$, so that $\gamma_{\mu\nu}a^\mu b^\nu=-({\vec a}\cdot{\vec b})$. We use powers of the inverse of the speed of light, $c^{-1}$, and the gravitational constant, $G$ as bookkeeping devices for order terms: in the low-velocity ($v\ll c$), weak-field ($GM/r\ll c^2$) approximation, a quantity of ${\cal O}(c^{-2})\simeq{\cal O}(G)$, for instance, has a magnitude comparable to $v^2/c^2$ or $GM/c^2r$. The notation ${\cal O}(a^k,b^\ell)$ is used to indicate that the preceding expression is free of terms containing powers of $a$ greater than or equal to $k$, and powers of $b$ greater than or equal to $\ell$.},
\begin{eqnarray}
g_{00}&=& 1-\frac{2}{c^2}w+\frac{2}{c^4}w^2+{\cal O}(c^{-6}), \quad
g_{0\alpha}= -\gamma_{\alpha\lambda}\frac{4}{c^3}w^\lambda+{\cal O}(c^{-5}),\quad
g_{\alpha\beta}= \gamma_{\alpha\beta}+\gamma_{\alpha\beta}\frac{2}{c^2} w+{\cal O}(c^{-4}).\label{eq:gab}
\end{eqnarray}

For the standard non-rotating GCRS, the scalar gravitational potential is formed as a linear superposition of the gravitational potential $U_{\rm E}$ of the isolated Earth, and the tidal potential $u_{\rm E}^{\rm tidal}$ of all other solar system bodies:
\begin{eqnarray}
w_{\rm E}&=&U_{\rm E}+ u^{\tt tidal}_{\rm E}+{\cal O}(c^{-4}).~~~
\label{eq:pot_loc-w_0}
\end{eqnarray}

$U_{\rm E}$ is conveniently represented in the form
\begin{eqnarray}
U_{\rm E}({\vec x})&=&
\frac{GM_{\rm E}}{r_{\rm }}\Big(1-\sum_{\ell=2}^\infty\Big(\frac{R_{\rm E}}{r_{\rm }}\Big)^\ell J_\ell P_{\ell0}(\cos\theta)+\sum_{\ell=2}^\infty\sum_{k=1}^{+\ell}\Big(\frac{R_{\rm E}}{r}\Big)^\ell P_{\ell k}(\cos\theta)(C_{\ell k}\cos k\phi+S_{\ell k}\sin k\phi)\Big),~~~~~
\label{eq:U-harm}
\end{eqnarray}
where $M_{\rm E}$ is the Earth's mass, $R_{\rm E}$ is its equatorial radius, $J_\ell=-C_{\ell0}$ are the zonal harmonics coefficients of the Earth mass distribution, $P_{\ell k}$ are the Legendre polynomials, while $C_{\ell k}$ and $S_{\ell k}$ are relativistic normalized spherical harmonic coefficients that characterize the Earth.

At the level of accuracy required by the ACES experiment, the tidal potential $u^{\rm tidal}_{\rm E}$ can be given in terms of Newtonian contributions only (mostly due the Moon and the Sun):
{}
\begin{eqnarray}
u^{\tt tidal}_{\rm E}&=&\sum_{b\not={\rm E}}\Big(U_b({\vec r}^{}_{b{\rm E}}+\vec{x})-U_b({\vec r}^{}_{b{\rm E}}) - \vec{x}\cdot {\vec \nabla} U_b ({\vec r}^{}_{b{\rm E}})\Big)\simeq\sum_{b\not={\rm E}}\frac{GM_b}{2r^3_{b{\rm E}}}\Big(3(\vec{n}^{}_{b{\rm E}}\cdot\vec{x})^2-\vec{x}^2\Big)+
{\cal O}(r^{-4}_{b{\rm E}},c^{-2}),
\label{eq:u-tidal-E}
\end{eqnarray}
where $U_b$ is the Newtonian gravitational potential of body $b$, $\vec{r}_{b{\rm E}}$ is the vector connecting the center of mass of body $b$ with that of the Earth, and $\vec \nabla U_b$ denotes the gradient of the potential. The potential can be expanded to higher order terms if necessary.

The vector harmonic  potential $w^\lambda_{\rm E}$ captures the contribution of the Earth's rotation, defined as:
{}
\begin{eqnarray}
w^\lambda_{\rm E}
&=&
-\frac{GM_{\rm E}}{2r^3}[{\vec x}\times{\vec S}_{\rm E}]^\lambda+{\cal O}(r^{-4}, c^{-2}).
\label{eq:pot_loc-w_a+}
\end{eqnarray}
Eq.~(\ref{eq:pot_loc-w_a+}) explicitly accounts only for the largest rotational moment,  ${\vec S}_{\rm E}$, which is the Earth's spin moment (angular momentum per unit of mass),  ${\vec S}_{\rm E}\simeq 9.8\times 10^8~{\rm m}^2$/s.

Although ACES data will be transmitted, received, and analyzed in GCRS, using TCRS, SCRS, and ACRS, we still need access to a global inertial reference frame to account for the presence of the solar system bodies. For these purposes, we need one {\it global} coordinate chart, defined for the inertial reference frame that covers the entire solar system. The Solar System Barycentric Coordinate Reference System (BCRS) has its origin at the solar system barycenter. It is a convenient reference system for the purpose of describing the motion of the Earth, the Moon, and the Sun \cite{Turyshev-Toth:2013,Turyshev:2012nw,Turyshev:2014dea}, whose dynamics will be important for ACES. In particular, the ellipticity of Earth's orbit introduces annual variations of solar gravity potential at the location of the ACES clock which must be accounted for, as it was done, for instance, in \cite{Ashby-etal-2007}. Otherwise, GCRS fully satisfies the needs of the ACES experiment.

\subsection{Terrestrial reference frames}
\label{sec:GCRS}

The formulation of the GCRS treats the Earth's trajectory in the solar system as being determined solely by gravitational forces -- a well justified assumption. Indeed, the largest nongravitational force acting on the Earth is that due to the solar radiation pressure, which is responsible for acceleration of $1.95\times 10^{-16}~{\rm m/s}^2$, which is negligible for the ACES experiment. Further details on the formulation of the GCRS are in \cite{Turyshev:2012nw,Turyshev-Toth:2013,Turyshev:2014dea,Soffel:2003cr}.

In Ref.~\cite{Turyshev-Toth:2013}, we showed that the transformations between the harmonic coordinates of the GCRS $\{x^m\}\equiv(ct,{\vec x})$ and a non-rotating body-centric reference system $\{y^m_a\}\equiv(ct_a,{\vec y}_a)$ associated with a body $a$ may be written as below:
{}
\begin{eqnarray}
t&=& t_a+
c^{-2}\Big\{({\vec v}_{a_0}\cdot {\vec y}_a)+\int_{t_{a_0}}^{t_a}\!\!
\Big[{\textstyle\frac{1}{2}}{\vec v}^2_{a_0}
+\overline{w}^a_{\rm ext}
\Big]dt'_a\Big\}+{\cal O}(c^{-4})\,t_a,
\label{eq:trans++x0}\\
{\vec x}&=&{\vec x}^{}_{a_0}+ {\vec y}^{}_a+
c^{-2}\Big\{{\textstyle\frac{1}{2}}{\vec v}^{}_{a_0} ({\vec v}^{}_{a_0}\cdot{\vec y}^{}_a)-\overline{w}_{\rm ext}^a{\vec y}^{}_a+
{\textstyle\frac{1}{2}}{\vec a}^{}_{a_0}{y}_a^2-{\vec y}^{}_a({\vec y}^{}_a\cdot {\vec a}^{}_{a_0})
\Big\}+{\cal O}(c^{-4}),
\label{eq:trans++xa}
\end{eqnarray}
where ${\vec x}_{a_0}={\vec x}_{a_0}(t)$ is the vector that connects the origin of the $\{x^m\}$ reference system with that of the $\{y^m_a\}$.

The quantity $\overline{w}^a_{\rm ext}$ in (\ref{eq:trans++x0})--(\ref{eq:trans++xa}) is a combination of potentials  evaluated at the origin of a particular reference frame. This combination consists of
\begin{inparaenum}[i)]
\item the Newtonian gravitational potential (including, if necessary, multipole corrections) due to all bodies in the solar system other than body $a$, at the location of body $a$, and
\item a contribution from nongravitational forces acting on the body $a$. \end{inparaenum}
Next, ${\vec v}_{a_0}\equiv\dot{\vec x}_{a_0}$ is the geocentric velocity of the body $a$ and ${\vec a}_{a_0}\equiv\ddot{\vec x}_{a_0}=-{\vec \nabla} w^a_{\rm ext}+{\cal O}(c^{-2})$ its Newtonian acceleration due to $w^a_{\rm ext}$, i.e., the combined effect from gravitational and nongravitational forces acting on the body.

Introducing ${\vec r}_{a}={\vec x}-{\vec x}_{a_0}$, the inverses of the transformations (\ref{eq:trans++x0})--(\ref{eq:trans++xa}) can be written as \cite{Turyshev:2012nw,Turyshev-Toth:2013,Turyshev:2014dea,Turyshev-etal:2012}
{}
\begin{eqnarray}
t_a&=& t-c^{-2}\Big\{({\vec v}^{}_{a_0}\cdot {\vec r}^{}_a)+\!\int_{t_{0}}^{t}\!\Big[{\textstyle\frac{1}{2}}{\vec v}^2_{a_0}+
\overline{w}_{\rm ext}^a \Big]dt'\Big\}+{\cal O}({c^{-4}})\,t,
\label{eq:trans++y0}\\
{\vec y}_a&=& {\vec r}_a+
c^{-2}\Big\{{\textstyle\frac{1}{2}}{\vec v}_{a_0} ({\vec v}_{a_0}\cdot{\vec r}_a)+\overline{w}_{\rm ext}^a{\vec r}_a
-{\textstyle\frac{1}{2}}{\vec a}^{}_{a_0}{r}^2_a+{\vec r}^{}_a ({\vec r}{}_a\cdot{\vec a}_{a_0})
\Big\}+{\cal O}(c^{-4}).
\label{eq:trans++ya}
\end{eqnarray}
Note that the $c^{-4}$ terms in (\ref{eq:trans++x0}), (\ref{eq:trans++y0}) are of the order of $\sim v^4_{\rm ISS}/c^4\simeq 4.3\times 10^{-19}$ and are therefore negligible for the ACES experiment. For complete post-Newtonian form of transformations (\ref{eq:trans++x0})--(\ref{eq:trans++ya}), consult Ref.~\cite{Turyshev-Toth:2013}.

Applying the coordinate transformations (\ref{eq:trans++x0})--(\ref{eq:trans++xa}) together with the external potentials $w^a_{\rm ext}$ and acceleration ${\vec a}_a$, from the metric tensor of the GCRS given by (\ref{eq:gab}) one can derive the metric tensor $g^{\rm a}_{mn}$ of the  non-rotating  coordinate reference system associated with an object $a$ (see details in \cite{Turyshev-Toth:2013}):
{}
\begin{eqnarray}
g^{\rm a}_{00}&=& 1-\frac{2}{c^2}w_{\rm a}+\frac{2}{c^4}w^2_{\rm a}+{\cal O}(c^{-6}), \quad
g^{\rm a}_{0\alpha}= -\gamma_{\alpha\lambda}\frac{4}{c^3}w^\lambda_{\rm a}+{\cal O}(c^{-5}),\quad
g^{\rm a}_{\alpha\beta}= \gamma_{\alpha\beta}+\gamma_{\alpha\beta}\frac{2}{c^2} w_{\rm a}+{\cal O}(c^{-4}),~~~
\label{eq:gab-harm}
\end{eqnarray}
where $w_{\rm a}$ and $w^\lambda_{\rm a}$ are the scalar and vector harmonic potentials representing gravity and inertia in a particular coordinate reference system associated with the body $a$.

For completeness, we also include the TCRS, which can be obtained by transforming the GCRS metric $g_{mn}^{\rm E}$ using (\ref{eq:trans++x0})--(\ref{eq:trans++xa}), where the ``external'' potential ${w}_{\rm ext}^{\rm C}$ is now the gravitational potential $w^{}_{\rm E}$ given by (\ref{eq:pot_loc-w_0}) and evaluated at the surface of the Earth:
{}
\begin{eqnarray}
{w}_{\rm ext}^{\rm C}({\vec y}_{\rm C})&=&U_{\rm E}({\vec y}_{\rm C})+
\sum_{b\not={\rm E}}\Big(U_b({\vec r}^{}_{b{\rm E}}+\vec{y}^{}_{\rm C})-U_b({\vec r}^{}_{b{\rm E}}) - \vec{y}_{\rm C}\cdot {\vec \nabla} U_b ({\vec r}^{}_{b{\rm E}})\Big) +{\cal O}(c^{-2}),~~~
\label{eq:TCRS-w_0}
\end{eqnarray}
where ${\vec y}_{\rm C}$ is the position vector in the GCRS of a particular ground station. Note that $U_{\rm E}({\vec y}_{\rm C})$ must be treated as the potential of an extended body and include a multipolar expansion with sufficient accuracy, taking into account time-dependent terms due to tidal effects on the elastic Earth. The corresponding acceleration is $\vec{a}_{\rm C}=-\nabla{w}_{\rm ext}^{\rm C}({\vec y}_{\rm C})$.

The proper time $\tau_{\rm C}$, kept by a clock $C$ located at the GCRS coordinate position ${\vec r}_{\rm C}(t)$, and moving with the coordinate velocity ${\vec v}_{\rm C} = d{\vec r}_{\rm C}/dt=[{\boldsymbol\omega}^{}_{\rm E}\times{\vec r}_{\rm C}]$, where ${\boldsymbol\omega}_{\rm E}$ is the angular rotational velocity of the Earth at $C$, is determined by
\begin{equation}
\frac{d\tau_{\rm C}}{dt} = 1 -\frac{1}{c^2}\Big[
{\textstyle\frac{1}{2}}\omega_{\rm E}^2 r^2_{\rm C}(\theta)\sin^2\theta
+ U_{\rm E}({\vec r}_{\rm C})\Big]+{\cal O}(3.89 \times 10^{-17}),
\label{eq:proper-coord-t+}
\end{equation}
where $\theta$ is the latitude of $C$ and the error bound is set by the lunar tides at the location of the tracking station.

The largest contribution to $d\tau_{\rm C}/dt$ comes from the velocity and mass monopole terms, which are estimated to produce an effect of the order of $c^{-2}({\textstyle\frac{1}{2}}\omega_{\rm E}^2 R^2_{\rm E}+G M_{\rm E}/R_{\rm E})\sim 6.97\times 10^{-10}$. The quadrupole term produces contribution of the order of $c^{-2}G M_{\rm E}J_2/(2R_{\rm E})\sim 3.77\times 10^{-13}$, which is large enough to be included in the model. Contributions of other zonal harmonics ranging from $-c^{-2}3G M_{\rm E}J_4/(8R_{\rm E})\sim 4.23\times 10^{-16}$ (from $J_4$) to $c^{-2}5G M_{\rm E}J_6/(16R_{\rm E})\sim 1.18\times 10^{-16}$ (from $J_6$). Although individual contributions of these and other terms are quite small to warrant their place in the model, their cumulative effect may be noticeable at the level of up to $3\times 10^{-15}$.

In practice, time measurements are based on averages of clock and frequency measurements on the Earth surface \cite{Moyer:2003}. For this purpose, the time coordinate called Terrestrial Time (TT) is defined. TT is related to TCG$=t$ linearly by definition:
\begin{equation}
\frac{dt_{\rm TT}}{dt}=1-L_{\rm G}.
\end{equation}
IAU Resolution B1.9 (2000) turned $L_G$ into a defining constant with its value fixed to $L_{\rm G}=6.969290134\times 10^{-10}$.

This definition accounts for the secular term due to the Earth's potential when converting between TCG and the time measured by an idealized clock on the Earth geoid \cite{Moyer:2003,Klioner:2008,Klioner-etal:2010,Kopeikin-book-2011}. Using Eq.~(\ref{eq:proper-coord-t+}), we also have
\begin{equation}
\frac{d\tau_{\rm C}}{dt_{\rm TT}}=\frac{d\tau_{\rm C}}{dt}\frac{dt}{dt_{\rm TT}}=1+L_{\rm G}-\frac{1}{c^2}\Big[{\textstyle\frac{1}{2}}{\vec v}^2_{\rm C}+ U_{\rm E}({\vec y}_{\rm C})\Big]+{\cal O}(3.89 \times 10^{-17}).
\label{eq:t-TT}
\end{equation}

\subsection{Spacecraft motion in the GCRS}
\label{sec:eqm-GCRS}

In the GCRS, the relativistic equations of motion of an artificial Earth satellite are given as \cite{Moyer:2003,MG2005,Turyshev:2012nw,Turyshev:2014dea}:
{}
\begin{eqnarray}
\ddot{\bf r}_{\rm }&=&-
\frac{GM_{\rm E}}{r_{\rm }^3}{\vec r}_{\rm }
-\frac{\partial}{\partial {\vec r}_{\rm }}\Big\{\frac{GM_{\rm E}}{r}\sum_{\ell=2}^\infty\sum_{k=0}^{+\ell}\Big(\frac{R_{\rm E}}{r}\Big)^\ell P_{\ell k}(\cos\theta)(C_{\ell k}\cos k\phi+S_{\ell k}\sin k\phi)\Big\}+\nonumber\\
&&\hskip 0pt +\,
\frac{GM_{\rm E}}{c^2r_{\rm }^3}\Big\{\Big[2(\beta+\gamma)\frac{GM_{\rm E}}{r_{\rm }}-\gamma \dot{\vec r}_{\rm }\cdot \dot{\vec r}_{\rm }\Big]{\bf r}_{\rm }+2(\gamma+1) ({\vec r}_{\rm }\cdot \dot{\vec r}_{\rm })\dot{\vec r}_{\rm }\Big\}+\nonumber\\
&&\hskip 0pt +\,
(\gamma+1)\frac{GM_{\rm E}}{c^2r_{\rm A}^3}
\Big\{\big[\dot{\vec r}_{\rm }\times\big({\vec S}_{\rm E}-3{\vec n}_{\rm }({\vec n}_{\rm }\cdot {\vec S}_{\rm E})\big)\big]\Big\}+
(2\gamma+1)\frac{GM_{\rm S}}{c^2\rho^3}
\Big\{\big[[{\vec \rho}\times \dot{\vec \rho}]\times\dot{\vec r}_{\rm }\big]\Big\}-\nonumber\\
&&\hskip 0pt -\,
\sum_{b\not={\rm E}}\frac{GM_b}{r^3_{b{\rm E}}}\Big(3(\vec{n}^{}_{b{\rm E}}\cdot\vec{r}_{\rm })\vec{n}^{}_{b{\rm E}}-\vec{r}_{\rm }\Big)+{\vec a}_{\rm NG}+
{\cal O} (10^{-12}~{\rm m/s}^2),
\label{eq:eqm-GEO}
\end{eqnarray}
where ${\vec r}_{\rm }$ is the position of the satellite in the GCRS, $\dot{\vec r}$ its velocity, $\vec{n}^{}_{\rm }=\vec{r}^{}_{\rm }/{r}^{}_{\rm }$ is the unit vector in the direction of the spacecraft, $\vec\rho$ is the Sun-Earth vector, ${\vec S}_{\rm E}$ is the Earth's angular momentum per unit mass, and $GM_{\rm E}$ and $GM_{\rm S}$ are the gravitational coefficients of the Earth and Sun, respectively. Furthermore, ${\vec r}_{b{\rm E}}={\vec r}_{\rm E}-{\vec r}_b$ is the position of the body $b$ with respect to the Earth in the BCRS and ${\vec n}_{b{\rm E}}={\vec r}_{b{\rm E}}/{r}_{b{\rm E}}$ is the unit vector in this direction. The first two terms in this equation represent the gradient of the Newtonian relativistic gravitational potential of the extended Earth, given by (\ref{eq:U-harm}). The coefficients $\beta$ and $\gamma$ are the Parameterized Post-Newtonian (PPN) parameters, equal to 1 in general relativity. Lastly, ${\vec a}_{\rm NG}$, is the contribution of nongravitational forces (to be discussed in Sec.~\ref{sec:ng-forces}).

As expected, Newtonian acceleration has the largest effect on Earth-orbiting spacecraft. For satellites in the vicinity of the Earth (up to GEO orbit) the terms in Eq.~(\ref{eq:eqm-GEO}) can be evaluated with respect to the main Newtonian acceleration, as follows: The Schwarzschild terms (second line) are a few parts in $10^{10}$ (high orbits) to $10^9$ (low orbits) smaller; the effects of Lense-Thirring precession (frame-dragging, first term on the third line) and the geodesic (de Sitter) precession (last term on the third line) are about $10^{11}$ to $10^{12}$ smaller. The main effect of the Schwarzschild terms is a secular shift in the argument of perigee while the Lense-Thirring and de Sitter terms cause a precession of the orbital plane at a rate of the order of 0.8 mas/yr (geostationary) to 180 mas/yr (low orbit) for Lense-Thirring and 19 mas/yr (independent of orbital altitude) for de Sitter. The Lense-Thirring terms are less important than the geodesic terms for orbits higher than LAGEOS (altitude above 6,000 km) and more important for orbits lower than LAGEOS.

\subsubsection{Gravitational effects on the ISS orbit}

Eq.~(\ref{eq:eqm-GEO}) consists of seven terms, the first of which represents the largest force acting on the ISS, the contribution of the Earth gravitational monopole ($\sim 8.69~{\rm m/s}^2$). The second term arises from treating the Earth as an extended body, and represents the contribution from the spherical harmonics, contributing a few parts in $10^{-3}$ (for $J_2$) to a few parts in $10^{-6}-10^{-7}$ (for harmonics of higher order) relative to the monopole term. The third term is the post-Newtonian contribution from the Schwarzschild metric, contributing up to $2.85\times 10^{-9}$~m/s$^2$, including also the contribution due to the eccentricity of the ISS orbit ($e_{\rm ISS}\sim 0.0006$), which is $\sim 1.37\times 10^{-11}$~m/s$^2$. The fourth term represents the Lense-Thirring precession due to the Earth's angular momentum, ${\vec S}_{\rm E}\simeq 9.8\times 10^8~{\rm m}^2$/s, amounting to an acceleration of the order of $\sim 2.15\times 10^{-10}$~m/s$^2$. The fifth term is due to the contribution of solar gravity on the ISS orbit, with a magnitude of $\sim 4.56\times 10^{-11}$~m/s$^2$. The sixth term represents the tidal potential (mostly due to the Earth and the Sun). The tidal acceleration due to the moon is of the order of $\sim 5.85\times 10^{-7}$~m/s$^2$ whereas for the sun it is $\sim 2.69\times 10^{-7}$~m/s$^2$. Finally, nongravitational forces are captured by the symbol $\vec{a}_{\rm NG}$, and will be discussed in the next subsection.

The relativistic geocentric equations of motion of a satellite that are recommended by the IERS \cite{Petit-Luzum:2010} may include the contribution from the relativistic quadrupole moment of the Earth at the $1/c^2$ order. It is estimated that the corresponding $J_2$-term would contribute $\sim 2.85\times 10^{-11}~{\rm m/s}^2$ for the ISS. As such, the relativistic effects of the Earth's oblateness are small for the ISS and may be neglected in the analysis.

The independent variable of the satellite equations of motion may be, depending on the time transformation being used, either TT or TCG. Although the distinction is not essential to compute this relativistic correction, it is important to account for it properly in the Newtonian part of the acceleration.

\subsubsection{Non-gravitational forces acting on the ISS}
\label{sec:ng-forces}

The trajectories of Earth-orbiting spacecraft can be significantly affected by nongravitational forces. Atmospheric drag is especially significant for the ISS, which has a very large surface area compared to its weight (i.e., it has a low ballistic coefficient), while moving in a low Earth orbit. Another significant contribution is due to solar radiation pressure and related torques. Other, lesser contributions include radiation pressure from the Earth, interactions with the Earth's magnetic fields, outgassing, thermo-elastic deformations and extended structure vibrations of the ISS \cite{Sosnica:2015}. However, for the ISS these contributions are small compared to the effect of the main nongravitational forces and will be addressed elsewhere.

The ISS is orbiting at an altitude of $\sim 400$~km. It is constantly losing altitude due to atmospheric drag. The station is regularly boosted
to prevent atmospheric re-entry. The atmospheric drag force on the station acts in the opposite direction of its velocity vector and is responsible for deceleration given by
{}
\begin{equation}
\vec{a}_{\rm ad}=-\frac{c_dA}{2m}\rho_{\rm atm} v^2 {\vec n}_v,
 \label{eq:ad}
\end{equation}
where $c_d$ is the drag coefficient, $\rho_{\rm atm}$ is the air density at
the station location, $v$ is the station's velocity with respect to the atmosphere, ${\vec n}_v={\vec v}/v$ is the unit vector in the direction of velocity, and $A$ is the cross-sectional area of the station in the direction of its motion through the atmosphere. The drag coefficient and cross-sectional area are dependent on the geometric shape and orientation of the station. These are generally determined through experiment or numerical simulations. Earth orbiting satellites typically have very high drag coefficients $c_d$ in the range of about 2 to 4.

The atmospheric density $\rho_{\rm atm}$ varies by up to two orders of magnitude as it responds to solar and geomagnetic activity \cite{Sosnica:2015}. The MSISE-90 model of the Earth's upper atmosphere\footnote{For details on the MSISE-90 model of the Earth's upper atmosphere, please consult {\tt http://www.braeunig.us/space/atmos.htm}}  provides $\rho_{\rm atm}$ as a function of solar activity.  At the altitude of $400$~km, the atmospheric density is expected to be in the range $\rho_{\rm atm}=(3.89-50.4)\times 10^{-12}~{\rm kg/m}^3$, corresponding to mean to extreme high solar activity.

To estimate the magnitude of $\vec{a}_{\rm ad}$, we use the current mass of the ISS\footnote{For data on the ISS, see {\tt http://www.nasa.gov/mission$\_$pages/station/main/onthestation/facts$\_$and$\_$figures.html}} $m=m_{\rm ISS}=4.2\times 10^5$~kg, see Table~\ref{tb:params}. The largest cross-sectional area, based on the size of the solar panels, can be estimated at $A\lesssim 8\times(36.5~{\rm m}\times 11.6~{\rm m})=3,387 ~{\rm m}^2$. We use $c_d=2$ for the ISS used in orbit analysis\footnote{See details at \url{http://spaceflight.nasa.gov/realdata/sightings/SSapplications/Post/JavaSSOP/orbit/ISS/SVPOST.html}}. Substituting these values in Eq.~(\ref{eq:ad}), we obtain
\begin{equation}
|\vec{a}_{\rm ad}| \simeq (1.8-23.9)\times 10^{-6}~{\rm  m}/{\rm s}^2,
\label{eq:acc_ad}
\end{equation}
These values\footnote{Our estimates are consistent with the values derived in \cite{Montenbruck-etal:2011} that relied on the data representing earlier configuration of the ISS.} (corresponding to mean and extreme high solar activity) are approximately three orders of magnitude larger than relativistic accelerations in the motion of the ISS.

Computational models for solar radiation pressure are usually developed prior to launch. These models can be used to compute the acceleration as a function of spacecraft orientation and solar distance:
\begin{equation}
\vec{a}_{\rm sr}(r)= \frac{2 f_\odot A}{cmr^2}\vec n_\odot,
 \label{eq:sr}
\end{equation}
where $f_\odot=1,367 ~{\rm W/(m}^{2}{\rm AU}^{2})$ is the solar constant at 1 AU, $\vec n_\odot$ is the unit vector in the direction of the Sun and $r$ is the distance between the spacecraft and the Sun. Using the physical parameters of the ISS and its orbit, an upper limit of Eq.~(\ref{eq:sr}) can be obtained:
\begin{equation}
|\vec{a}_{\rm sr}| \lesssim 7.3\times 10^{-8}~{\rm  m}/{\rm s}^2,
\label{eq:sr-acc}
\end{equation}
which is nearly $10^8$ times smaller than the Newtonian acceleration and over $\sim10^2\text{ to }10^3$ times smaller than the deceleration due to atmospheric drag. However, solar radiation pressure will in general perturb the eccentricity, which may have to be counteracted by actively controlling the orbit. This acceleration, if not accounted for, will result in an error in the estimated radial position of the ISS. Over a time interval $\Delta t$ this error is $\delta r = {\textstyle\frac{1}{2}}\delta a \,\Delta t^2$, where  ${\delta a}$ is the unmodeled acceleration. If the effect of solar radiation pressure was completely unmodeled, over a period of $\Delta t=10^3$~s, the position error due to solar pressure would be $\sim$0.04~m, which is negligible over the time scale of one orbital revolution.

\subsubsection{Equation of motion appropriate for the ISS}

The orbit of the ISS (hence, ACES) is presumed known to an accuracy of $\sim$10~m \cite{Duchayne-etal:2009,Montenbruck-etal:2011}. Atmospheric drag corrupts the orbital estimates for the ISS, making it impractical to keep terms below $\sim 1\times 10^{-7}~{\rm  m}/{\rm s}^2$. Consequently, we may neglect all the relativistic terms (${\cal O}(c^{-2})$ and smaller), i.e., terms three through five in (\ref{eq:eqm-GEO}), truncating the equation of motion to its Newtonian form:
{}
\begin{eqnarray}
\ddot{\bf r}&=&-
\frac{GM_{\rm E}}{r^3}{\vec r}
-\frac{\partial}{\partial {\vec r}}\Big\{\frac{GM_{\rm E}}{r}\sum_{\ell=2}^\infty\sum_{k=0}^{+\ell}\Big(\frac{R_{\rm E}}{r}\Big)^\ell P_{\ell k}(\cos\theta)(C_{\ell k}\cos k\phi+S_{\ell k}\sin k\phi)\Big\}-
\nonumber\\
&&\hskip 0pt -\,
\sum_{b\not={\rm E}}\frac{GM_b}{r^3_{b{\rm E}}}\Big(3(\vec{n}^{}_{b{\rm E}}\cdot\vec{r}_{\rm })\vec{n}^{}_{b{\rm E}}-\vec{r}_{\rm }\Big)-
\frac{c_dA}{2m}\rho_{\rm atm} |\dot{\vec r}|^2 {\vec n}_v+
{\cal O} (1\times10^{-7}~{\rm m/s}^2),~~~
\label{eq:eqm-ISS-real}
\end{eqnarray}
where the bound is provided by the solar radiation pressure (\ref{eq:sr-acc}).

\subsection{Light travel time}
\label{sec:effects_on_ltt}

Knowing the time that it takes for a signal to travel between the Earth and a spacecraft is required both for radio-metric navigation and for clock synchronization. Given a time of transmission $t_1$ and a time of reception $t_2$, the positions of the transmitter ($\vec{x}_{\rm A}$) and receiver ($\vec{x}_{\rm B}$) are related by the light time equation
{}
\begin{eqnarray}
t_2-t_1&=&\frac{1}{c}{\cal R}\big({\vec x}_{\rm A}\big(t_1\big),{\vec x}_{\rm B}(t_{\rm 2})\big),
\label{eq:light-time}
\end{eqnarray}
where ${\cal R}\big({\vec x}_{\rm A},{\vec x}_{\rm B}\big)$ is the total geodesic one-way distance traveled by light between the transmitter and receiver. To first order in $G$, in Ref.~\cite{Turyshev:2014dea} it was derived as
{}
\begin{eqnarray}
{\cal R}\big({\vec x}_{\rm A}\big(t_1\big),{\vec x}_{\rm B}(t_{\rm 2})\big)= |{\vec x}_{\rm B}(t_{\rm 2}) - {\vec x}_{\rm A}(t_{\rm 1})|+(1+\gamma)\frac{G M_{\rm E}}{c^2}
\ln\left[\frac{r_{\rm A1}+r_{\rm B2}+r_{\rm A1B2}}{r_{\rm A1}+r_{\rm B2}-r_{\rm A1B2}}\right]+{\cal O}\big(c^{-3}, J_2\big),
\label{eq:R-total}
\end{eqnarray}
with $r_{\rm A1}=|{\vec x}_{\rm A}(t_1)|$ and $r_{\rm B2}=|{\vec x}_{\rm B}(t_2)|$ being the distances of the transmitter and receiver from the origin of GCRS and $r_{\rm A1B2}=|{\vec x}_{\rm B}(t_{\rm 2}) - {\vec x}_{\rm A}(t_{\rm 1})|$ is their spatial separation. The logarithmic contribution in (\ref{eq:R-total}) is the Shapiro gravitational time delay \cite{Moyer:2003} that, in the case of ACES, is all due to the Earth (contributions from the Moon and the Sun are negligible). Between the ISS and the ground, the Shapiro time delay is 0.5--3.1~mm (i.e., 1.8--10.4~ps).
The next order term in this expression would be the one due to Earth's quadrupole \cite{Turyshev:2014dea} whose presence extends the standard formulation given in  \cite{Moyer:2003,Turyshev:2008}. The largest contribution from the quadruple will be for a ground-based receiver. It amounts to
$\sim3~\mu$m (or 0.01~ps) and is negligible for ACES, providing the corresponding bound in (\ref{eq:R-total}).

These computations are applicable when signals travel in a vacuum. This is not the case for signals between ground stations and spacecraft in low-Earth orbit. These signals experience propagation details due to the charged particle environment in the exosphere and, most notably, due to the properties and composition of the troposhere. There is extensive literature available on this topic (see., e.g., \cite{Abshire-Gardner:1985,Gardner:1976,Laurent-etal:2008,Delva-etal:2012,Mendes-etal:2002,Mendes-Pavlis:2004,Fridelance-etal:1997,Bender:1992}), a comprehensive discussion of which is beyond the scope of our present paper.

\subsection{Coordinate systems on the ISS}
\label{sec:crs-ISS}

Observables on board the ISS are naturally described in a coordinate system (the SCRS) associated with the space station's center of mass. However, this choice must be scrutinized for a precision experiment that is intended to detect minute gravitational effects, if the experiment is not located at the station's center of mass, as it is subject to tidal accelerations as well as nongravitational forces, such as structure vibrations. The ACES experiment is some distance away. For this reason, we also introduce the ACES Coordinate Reference System (ACRS), and use the coordinate transformations between the SCRS and the ACRS to estimate the relativistic corrections due to the location of ACES. We begin, though, by establishing the relationship between terrestrial and station reference frames.

\subsubsection{Coordinate transformations between GCRS and SCRS}
\label{sec:gscs-scrs}

To determine the metric tensor for the SCRS, we perform the coordinate transformation between the GCRS coordinates, $\{x^m_{\rm E}\}\equiv(x^0=ct, {\vec x})$, and coordinates of the non-rotating ISS-centric reference system $\{y^m_{\rm ISS}\}\equiv(ct_{\rm ISS},{\vec y}_{\rm})$ associated with the center of mass of the ISS. With the help of (\ref{eq:trans++x0})--(\ref{eq:trans++xa}), these transformations are taking the form:
\begin{eqnarray}
t&=&t_{\rm ISS}+c^{-2}\Big\{({\vec v}_{\rm ISS}\cdot {\vec y}_{\rm})+\int_{t_{{\rm ISS}_0}}^{t_{\rm ISS}}\Big[{\textstyle\frac{1}{2}}{\vec v}_{\rm ISS}^2+\overline{w}_{\rm ext}^{\rm ISS}\Big]dt'_{\rm ISS}\Big\}+{\cal O}(c^{-4})t_{\rm ISS},\label{eq:x-y_trans_0}
\\
{\vec x}&=&{\vec x}_{\rm ISS}+{\vec y}_{\rm}+
c^{-2}\Big\{{\textstyle\frac{1}{2}}{\vec v}^{}_{\rm ISS} ({\vec v}^{}_{\rm ISS}\cdot{\vec y}^{}_{\rm})-\overline{w}_{\rm ext}^{\rm ISS}{\vec y}^{}_{\rm}+
{\textstyle\frac{1}{2}}{\vec a}^{}_{\rm ISS}{y}_{\rm}^2-{\vec y}^{}_{\rm}({\vec y}^{}_{\rm}\cdot {\vec a}^{}_{\rm ISS})\Big\}+{\cal O}(c^{-4}),
\label{eq:x-y_trans_a}
\end{eqnarray}
where ${\vec x}_{\rm ISS}={\vec x}_{\rm ISS}(t)$ is the geocentric position vector of the ISS
and ${\vec v}_{\rm ISS}\equiv\dot{\vec x}_{\rm ISS}$ its geocentric velocity. The geocentric acceleration of the ISS, ${\vec a}_{\rm ISS}\equiv\ddot{\vec x}_{\rm ISS}$, is given by Eq.~(\ref{eq:eqm-ISS-real}), which is accurate to $\sim 1\times 10^{-7}~{\rm  m}/{\rm s}^2$ and accounts for Newtonian gravity of the Earth together with the contribution from the atmospheric drag.

\begin{table}[t!]
\vskip-15pt
\caption{Select parameters of the ACES mission \cite{Gomez:2010}, along with corresponding symbols and approximate formulae used in the text.
\label{tb:params}}
\begin{center}
\begin{tabular}{|l|c|c|}\hline
Parameter  &Symbol  &Value\\\hline\hline
ISS orbital altitude
&$h_{\rm ISS}$ &\phantom{0}400~km\phantom{/s}\\
ISS orbital eccentricity&$e_{\rm }$ &\phantom{000}0.0006\phantom{/s}\\
ISS orbital inclination&$i_{\rm ISS}$ &\phantom{}51.6$^\circ$\phantom{}\\
ISS mass&$m_{\rm ISS}$ &\phantom{}420,000~kg\phantom{}\\
ISS cross-sectional area\tablenote{Based on the size of the solar panels.} &$A_{\rm ISS}$ &\phantom{}3,400~m$^2$\phantom{}\\
ACES position in SCRS& $ y_{\rm A0}$ &\phantom{}30~m\phantom{}\\
Geocentric velocity      & $v_{\rm A0} = ({G M_{\rm E}/(R_\oplus+h_{\rm ISS})})^{1/2} $ &\phantom{.}7.67~km/s\phantom{}\\
Mean orbital frequency & $\omega_{\rm G}=({G M_{\rm E}/(R_\oplus+h_{\rm ISS})^3})^{1/2}$ & \phantom{0}1.13~mHz\phantom{k}\\
Geocentric acceleration & $a_{\rm A0} =G M_{\rm E}/(R_\oplus+h_{\rm ISS})^2$~\, &\phantom{0.}8.70~m/s$^2$\phantom{k}\\
\hline
\end{tabular}
\end{center}
\end{table}
\vskip-0pt

The ``external'' potential, ${w}_{\rm ext}^{\rm ISS}$, defined in the vicinity of the ISS is given as:
{}
\begin{eqnarray}
{w}_{\rm ext}^{\rm ISS}&=&U_{\rm E}({\vec y}_{\rm})+
\sum_{b\not={\rm E}}\Big(U_b({\vec r}^{}_{b{\rm E}}+\vec{y}^{}_{\rm})-U_b({\vec r}^{}_{b{\rm E}}) - {\vec{y}}^{}_{\rm }\cdot{\vec \nabla} U_b ({\vec r}^{}_{b{\rm E}})\Big)+({\vec a}_{\rm NG}\cdot{\vec y}_{\rm })
+{\cal O}(c^{-2}),~~~
\label{eq:SCRS-w_00}
\end{eqnarray}
with the last term being the contribution of the nongravitational forces acting on the space station. Evaluated at the origin of the SCRS ($\vec{y}=0$) this potential is:
{}
\begin{eqnarray}
\overline{w}_{\rm ext}^{\rm ISS}&=&\overline{U}_{\rm E}+
\sum_{b\not={\rm E}}\frac{GM_b}{2r^3_{b{\rm E}}}\Big(3(\vec{n}_{b{\rm E}}\cdot\vec{x}_{\rm ISS})^2-\vec{x}_{\rm ISS}^2\Big)+{\cal O}(c^{-2}),~~~
\label{eq:SCRS-w_0}
\end{eqnarray}
where $\overline{U}_{\rm E}$ is the contribution of extended Earth evaluated at the center of mass of the ISS, as given by Eq.~(\ref{eq:U-harm}) at $r=|{\vec x}_{\rm ISS}|=R_{\rm E}+h_{\rm ISS}$ (see Table~\ref{tb:params}).
We can now substitute (\ref{eq:U-harm}) in (\ref{eq:x-y_trans_0}) and evaluate all the terms. To facilitate our analysis, we present the resulting equation in differential form:
\begin{eqnarray}
\frac{dt_{\rm ISS}}{dt}&=&1-c^{-2}\Big\{{\textstyle\frac{1}{2}}{\vec v}_{\rm ISS}^2+\overline{U}_{\rm E}+\sum_{b\not={\rm E}}\frac{GM_b}{2r^3_{b{\rm E}}}\Big(3(\vec{n}_{b{\rm E}}\cdot\vec{x}_{\rm ISS})^2-\vec{x}_{\rm ISS}^2\Big)+({\vec a}_{\rm ISS}\cdot {\vec y}_{\rm})\Big\}+{\cal O}(c^{-4}).
\label{eq:t_ISS*}
\end{eqnarray}

The first term in curly braces in Eq.~(\ref{eq:t_ISS*}) is due to the orbital velocity of the ISS. The actual velocity of the ISS at the end of a time interval $\delta t$ may be written as the sum of the initial velocity $\vec{v}_{\rm ISS}(t)$ and perturbations from gravitational ($\delta\vec{v}_{\rm grav}(t)=\int_t^{t+\delta t}\vec{a}_{\rm grav}(t')dt'$) and nongravitational ($\delta\vec{v}_{\rm NG}(t)=\int_t^{t+\delta t}\vec{a}_{\rm NG}(t')dt'$) accelerations: $\vec{v}_{\rm ISS}(t+\delta t)=\vec{v}_{\rm ISS}(t)+\delta\vec{v}_{\rm grav}(t)+\delta\vec{v}_{\rm NG}(t)$.
Writing $\vec{v}_{\rm ISS_0}(t+\delta t)=\vec{v}_{\rm ISS}(t)+\delta\vec{v}_{\rm grav}(t)$ and replacing $\vec{a}_{\rm NG}(t)$ with a constant value $\vec{a}_{\rm NG}$ representing the maximum nongravitational acceleration allows us to estimate $|v_{\rm ISS}^2-v_{\rm ISS_0}^2|\lesssim 2(\vec{v}_{\rm ISS_0}\cdot\vec{a}_{\rm NG})\delta t+a^2_{\rm NG}\delta t^2$. The two largest terms in $\vec{a}_{\rm NG}$ are given by Eqs. (\ref{eq:acc_ad}) and (\ref{eq:sr}). Given $v_{\rm ISS_0}\sim 7.67~{\rm km}/{\rm s}$, over a time interval $\delta t=10^3~{\rm s}$, the contributions of these terms is given by, respectively, $2c^{-2}(\vec{v}_{\rm ISS_0}\cdot\vec{a}_{\rm ad})\delta t\lesssim 4.1\times 10^{-15}$ and $2c^{-2}(\vec{v}_{\rm ISS_0}\cdot\vec{a}_{\rm sr})\delta t\lesssim 1.3\times 10^{-17}$, whereas the contribution of the term $\propto a_{\rm NG}^2$ for that time interval is  $\sim{\cal O}(10^{-20})$. Therefore, we estimate that atmospheric drag contributes $\frac{1}{2}c^{-2}(\vec{v}_{\rm ISS_0}\cdot\vec{a}_{\rm ad})\lesssim 2.04\times 10^{-15}$ to clock desynchronization, whereas the contributions of other nongravitational accelerations are negligible.

The monopole term in the Newtonian potential of the Earth at the ISS location  $\overline{U}_{\rm E}$ is responsible for a contribution of $c^{-2}GM_{\rm E}/r_{\rm ISS}\simeq6.55\times10^{-10}$. We estimate that the quadrupole term produces a contribution of $c^{-2}G M_{\rm E}J_2R_{\rm E}^2/(2r^3_{\rm ISS})\simeq 3.14\times 10^{-13}$. Contributions of $J_4$ and $J_6$ are given by $-c^{-2}3G M_{\rm E}J_4R_{\rm E}^4/(8r^5_{\rm ISS})\sim 3.12\times 10^{-16}$ and  $J_6$ is $c^{-2}5G M_{\rm E}J_6R_{\rm E}^6/(16r^7_{\rm ISS})\sim 7.68\times 10^{-17}$, which are significant. Although individual contributions of other zonal harmonics (i.e, $J_k, k\geq 7$) are small, their cumulative effect may be noticeable at the level up to $\epsilon_{\rm ISS0}\approx3\times 10^{-16}$.

The constant rate $\epsilon_{\rm ISS0}$ would likely be absorbed in other terms during clock synchronization. What is important is the variability in  the entire error term $\epsilon_{\rm ISS}(t)=\epsilon_{\rm ISS0}+\delta\epsilon_{\rm ISS}(t)$, where the amplitude of the variable term $\delta\epsilon_{\rm ISS}(t)$ is due to seasonal changes in the Earth hydrosphere, crust, etc. and is expected to be of the order of  $\delta\epsilon_{\rm ISS}(t)\sim 3\times 10^{-17}$, resulting in the ultimate uncertainty in ACES clock instabilities $\delta({d\tau_{\rm A}}/{dt})_{\epsilon_{\rm ISS}}$ at that level. Thus, the mission may consider including the higher spherical harmonics in the model for Earth gravity potential. Conversely, the ACES data could be used for accurate determination and study of the lowest spherical harmonics (i.e, $J_k, k\in[2, 10]$), which could be a valuable scientific result for geodesy \cite{Svehla-etal:2008}.

The next $1/c^2$  term in (\ref{eq:t_ISS*}) is the sum of the Newtonian tides due to other bodies (mainly the Sun and the Moon) at the origin of the SRCS. These terms are small for the ISS being of the order of $4.40 \times 10^{-17}$ for the Moon and $2.02 \times 10^{-17}$ for the Sun, and, thus, they may be omitted for ACES.

The last $1/c^2$ term in (\ref{eq:t_ISS*}) is the acceleration-induced redshift for clocks on the ISS. Although this term vanishes at the origin of the SRCS, ACES will not be located at the center of mass of the ISS, but at some distance of ${y}_{\rm A0}=30$~m away from it. Such an offset leads to acceleration-induced redshift between a fictitious clock at the origin of the SGRS and the atomic clocks of the ACES package. The dominant term in $\vec{a}_{\rm ISS}$, given by $\vec{a}_{\rm ISS_0}$, yields a surprisingly large contribution: $c^{-2}({\vec a}_{\rm ISS_0}\cdot {\vec y}_{\rm A0}\big)\simeq2.91\times10^{-15}$. Contributions from nongravitational accelerations are ${\cal O}(10^{-20})$ and thus can be safely neglected.

As a result, the temporal part of coordinate transformation (\ref{eq:x-y_trans_0}) (also given by (\ref{eq:t_ISS*})) takes the following form:
\begin{eqnarray}
\frac{dt_{\rm ISS}}{dt}&=&1-c^{-2}\Big\{{\textstyle\frac{1}{2}}{\vec v}_{\rm ISS_0}^2+({\vec v}_{\rm ISS_0}\cdot{\vec v}_{\rm ad})+\overline{U}_{\rm E}+({\vec a}_{\rm ISS_0}\cdot {\vec y}_{\rm })\Big\}+{\cal O}(4.4\times 10^{-17}),
\label{eq:t_ISS}
\end{eqnarray}
where the accuracy bound is set by the omitted contribution from the lunar and solar tides (third $1/c^2$ term in (\ref{eq:t_ISS*})).

We also evaluate the $1/c^2$ terms present in the spatial transformations (\ref{eq:x-y_trans_a}). Taking, for example, the term with the Newtonian potential we determine that its contribution is of the order of $\propto c^{-2}\overline{U}_{\rm E} {y}_{\rm A0}\leq c^{-2}(GM_{\rm E}/r_{\rm ISS})y_{\rm A0}= 6.55\times 10^{-10}{y}_{\rm A0}$. Even for the largest separation on the ISS, $y_{\rm ISS}=120$~m, this term is only $\sim 7.86\times 10^{-8}$~m. As the uncertainty in the geocentric position of the ACES package is expected to be $\sim$1~m, this value is exceedingly small. The other $1/c^2$ terms in (\ref{eq:x-y_trans_a}) are also negligible. Thus, the entire package of $1/c^2$ terms in (\ref{eq:x-y_trans_a}) may be omitted.

As a result, the spatial part of the coordinate transformations between GCRS and SCRS (\ref{eq:x-y_trans_a}) takes a simple form:
{}
\begin{eqnarray}
{\vec x}&=&{\vec x}_{\rm ISS}+{\vec y}_{\rm}+
{\cal O}(6.55\times 10^{-10}){\vec y}_{\rm}.
\label{eq:x_ISS}
\end{eqnarray}

Compared to (\ref{eq:x-y_trans_0})--(\ref{eq:x-y_trans_a}), the coordinate transformations (\ref{eq:t_ISS})--(\ref{eq:x_ISS}) have a much simplified form, but are accurate enough to  describe the ACES observables and will be used them for this purpose in the reminder of the paper.

\subsubsection{Metric tensor of the SCRS on the ISS}

Using the formalism of Eq.~(\ref{eq:gab}), the metric tensor of the SCRS can be expressed in the harmonic coordinates of this SCRS, $\{y^m_{\rm ISS}\}\equiv(ct_{\rm ISS}, {\vec y}_{\rm})$, using scalar and vector potentials that incorporate both gravitational and nongravitational contributions \cite{Turyshev-etal:2012,Turyshev-Toth:2013}:
\begin{eqnarray}
w_{\rm ISS}&=&U_{\rm ISS}+ u^{\tt tidal}_{\rm ISS}+({\vec a}_{\rm NG}\cdot{\vec y}_{\rm})+{\cal O}(c^{-3}),~~~
\label{eq:pot_loc-w_ISS0} \\
{\vec w}_{\rm ISS}&=&-\frac{GM_{\rm E}}{2r^2_{\rm ISS}}[{\vec n}_{\rm ISS}\times{\vec S}_{\rm E}]-
{\textstyle\frac{1}{10}}\big(3{\vec y} ({\vec y}\cdot\dot{\vec a}^{}_{\rm NG})-\dot{\vec a}^{}_{\rm NG}{\vec y}^2\big)+
{\cal O}(c^{-2}).~~~
\label{eq:pot_loc-w_a_ISS0}
\end{eqnarray}

Of the terms present in Eq.~(\ref{eq:pot_loc-w_ISS0}), contributions to clock desynchronization by the gravitational potential due to the self-gravity of the ISS at a distance $y=|\vec{y}|=30$~m away from that point are given by $\delta(d\tau_{\rm ACES}/dt)_{\rm ISS}=c^{-2}U_{\rm ISS}=c^{-2}G m_{\rm ISS}/|y_{\rm ISS}|\simeq 1.0\times 10^{-23}$, which is negligible. Similarly, contributions from the tidal term, $\delta(d\tau_{\rm ACES}/dt)_{\rm tidal}=c^{-2}u^{\rm tidal}_{\rm ISS}=c^{-2}\big\{GM_{\rm E}[3(\vec{n}^{}_{\rm ISS}\cdot\vec{y}_{\rm })^2-\vec{y}^2_{\rm }]/2r^3_{\rm ISS}+ {\cal O}(y^{3}_{\rm },c^{-2})\big\}
\simeq 1.3\times 10^{-20}$, can also be dropped. Finally, estimating the contributions from nongravitational forces by using the largest nongravitational force, atmospheric drag, we find that the magnitude of these contributions, $\delta(d\tau_{\rm ACES}/dt)_{\rm NG}=c^{-2}({\vec a}_{\rm NG}\cdot{\vec y})\sim a_{\rm ad}y_{\rm A0}\simeq 8.2\times 10^{-21}$ means that the third term can also be omitted.

As the center of figure of the ISS does not coincide with its center of mass, the nongravitational forces may result in torques applied to the entire space station, changing the attitude orientation of the ISS. The ISS is maintained in a ``torque equilibrium attitude'' \cite{Kumar:1990}, which means that the cumulative effect of torque over a full orbit is near zero, but during an orbit, significant accelerations may be present. To estimate the order of magnitude of such torques, we model the space station as a spherical, homogeneous $m=4.2\times 10^5$~kg mass of $R=30$~m radius, with moment of inertia $I=2mR^2/5$, and assume that atmospheric drag, the largest nongravitational force acting on the station, is displaced by up to 1 m relative to the center-of-mass at some point during the orbit. The torque, then, is given by $\tau=\vec{r}\times(m\vec{a}_{\rm ad})\lesssim 10~{\rm N}\cdot{\rm m}$, and the rotational acceleration is $\alpha=\tau/I\sim 10^{-7}{\rm s}^{-2}$. At $y=30$~m, this translates into an acceleration of $a_{\rm trq}\sim 2.0\times 10^{-6}~{\rm m}/{\rm s}^2$, contributing $\delta(d\tau_{\rm ACES}/dt)_{\rm trq}\lesssim c^{-2}a_{\rm trq}y\sim 10^{-21}$, which is negligible.

The space station also rotates. Normally it completes one revolution per orbit, but higher rotation rates are also possible. However, even at the unrealistically high rotation rate of $\omega=1^\circ/{\rm s}$, $a_{\rm cf}=\omega^2_{\rm trq}y_{\rm A0}\simeq0.009~{\rm m/s}^2$, which would contribute to the frequency comparison of the ACES clocks at the level of $\delta (d\tau_{\rm ACES}/dt)_{\rm cf}=({\vec a}_{\rm trq}\cdot{\vec y}_{\rm A0})/c^2\simeq3.05\times 10^{-18}$, still small compared to our accuracy goal for ACES.

In Eq.~(\ref{eq:pot_loc-w_a_ISS0}), the first term contributes $c^{-4}GM_{\rm E}([{\vec n}_{\rm ISS}\times{\vec S}_{\rm E}]\cdot v_{\rm ISS})/(2r^2_{\rm ISS})\leq4\times 10^{-21}$, which is negligible. The second term, a gravitomagnetic term that results in the dragging of inertial frames and a clock redshift, is due to temporal variability of the nongravitational accelerations. Taking $\dot{a}^{}_{\rm NG}={a}^{}_{\rm ad}\omega_{\rm ISS}$, where $\omega_{\rm ISS}$ is the orbital frequency of the ISS, this contribution amounts to $c^{-4}(4 a_{\rm ad}\omega_{\rm ISS} y^2_{\rm A0}v_{\rm ISS})/5\leq1.9\times 10^{-35}$, which is exceedingly small.

As a result, the SCRS at the position of the ACES package on the ISS may be treated as inertial, with the metric tensor given by the Minkowski space-time with insignificant deviations due to non-inertial contributions:
{}
\begin{eqnarray}
g^{\rm ISS}_{00}&=& 1+{\cal O}(3.1\times 10^{-18})\equiv 1, \qquad
g^{\rm ISS}_{0\alpha}= {\cal O}(6.3\times 10^{-16})\equiv 0,\qquad
g^{\rm ISS}_{\alpha\beta}= \gamma_{\alpha\beta}+{\cal O}(3.1\times 10^{-18})\equiv \gamma_{\alpha\beta}.~~~
\label{eq:gab-ISS2}
\end{eqnarray}
This Minkowski metric covers the entire space station and may be used to analyze and process the data for the ACES experiment. In (\ref{eq:gab-ISS2}), the error terms for temporal, $g^{\rm ISS}_{00}$, and spatial, $g^{\rm ISS}_{\alpha\beta}$, components provided by a sudden nongravitational torques changing the attitude of the ISS by 1~${}^\circ/$s, while the error bound for the mixed terms, $g^{\rm ISS}_{0\alpha}$, is due to omitted contribution from the Earth's spin moment.
As a result, to a good approximation, the metric tensor of the SCRS can be taken in the form of the Minkowski metric (\ref{eq:gab-ISS2}), which is sufficient to describe the relativistic  observables of the ACES experiment. The SCRS may be treated as a locally inertial frame that covers the entire ISS. Together with the coordinate transformations (\ref{eq:t_ISS})--(\ref{eq:x_ISS}) the metric tensor (\ref{eq:gab-ISS2}) completes formulation of the SCRS.

\subsection{ACES Coordinate Reference System (ACRS)}
\label{sec:aces-cs}

To describe the observables of the ACES experiment, we need to introduce the ACRS coordinate reference system located at the center of mass of the ACES package. The coordinate transformation between the SCRS and the ACRS will be given in a form similar to \ref{eq:x-y_trans_0})--(\ref{eq:x-y_trans_a}):
\begin{eqnarray}
t_{\rm ISS}&=&t_{\rm A}+c^{-2}\Big\{({\vec v}_{\rm 0}\cdot {\vec y}_{\rm A})+\int_{t_{{\rm A}_0}}^{t_{\rm A}}\Big({\textstyle\frac{1}{2}}{\vec v}_{\rm 0}^2+\overline{w}^{\rm A}_{\rm ext}\Big)dt'_{\rm A}\Big\}+{\cal O}(c^{-4})t_{\rm A},
\label{eq:x-y_trans_0-AC*}\\
{\vec y}_{\rm ISS}&=&{\vec y}_{\rm A0}+{\vec y}_{\rm A}+
c^{-2}\Big\{{\textstyle\frac{1}{2}}{\vec v}^{}_{0} ({\vec v}^{}_{0}\cdot{\vec y}^{}_{\rm A})-\overline{w}_{\rm ext}^{\rm A}{\vec y}^{}_{\rm A}+
{\textstyle\frac{1}{2}}{\vec a}^{}_{0}{y}_{\rm A}^2-{\vec y}^{}_{\rm A}({\vec y}^{}_{\rm A}\cdot {\vec a}^{}_{0})\Big\}+{\cal O}(c^{-2}),
\label{eq:x-y_trans_a-AC*}
\end{eqnarray}
where ${\vec y}_{\rm A0}={\vec y}_{\rm A0}(t_{\rm ISS})$  is the positional vector of the ACES package with respect to the SCRS, with $|{\vec y}_{\rm A0}|=30$~m. Also, ${\vec v}^{}_{0}=\dot{\vec y}_{\rm A0}$ and ${\vec a}^{}_{0}=\ddot{\vec y}_{\rm A0}$ are the velocity and acceleration of the ACRS as seen from the SCRS, correspondingly. The ``external'' potential, $\overline{w}_{\rm ext}^{\rm A}$, is the potential due to the gravity of the ISS evaluated at the ACES location. Its magnitude is $\overline{w}_{\rm ext}^{\rm A}/c^2=U_{\rm ISS}/c^2\lesssim 1.0\times 10^{-23}$, making this term negligible.

The velocity ${\vec v}^{}_{0}=\dot{\vec y}_{\rm A0}$ is mostly due to extended structure vibrations of the ISS. If we assume that the ISS vibrates at the ACES location at $\omega_{\rm vib}=10$~Hz with the amplitude $\delta_{\rm vib}=0.1$~m, this motion leads to a vibrational velocity of $v_{\rm vib}=\delta_{\rm vib}\omega_{\rm vib}=1$~m/s, which will be responsible for a contribution of $\delta(dt_{\rm ISS}/dt_A)_{\rm vib}={\textstyle\frac{1}{2}}c^{-2}v_{\rm vib}^2=5.56\times 10^{-18}$ to ACES clock desynchronization. (As environmental factors on the ISS may increase the magnitude of this effect, it is important to analyze it in more detail, which we plan to do this in a separate publication.)

Based on this analysis, we conclude that the coordinate transformations between the SCRS and the ACRS (\ref{eq:x-y_trans_0-AC*})--(\ref{eq:x-y_trans_a-AC*}) may be given as
\begin{eqnarray}
t_{\rm ISS}&=&t_{\rm A}+c^{-2}({\vec v}_{\rm vib}\cdot {\vec y}_{\rm A})+{\cal O}(5.6\times 10^{-18})t_{\rm A},
\label{eq:x-y_trans_0-ACES}\\
{\vec y}_{\rm ISS}&=&{\vec y}_{\rm A0}+{\vec y}_{\rm A}+{\cal O}(5.6\times 10^{-18}){\vec y}_{\rm A},
\label{eq:x-y_trans_a-ACES}
\end{eqnarray}
where the bound is set by the vibrations of the ISS at the location of the ACES package. As a result, the metric tensor representing the ACRS has the nearly Minkowski form:
{}
\begin{eqnarray}
g^{\rm A}_{00}&=& 1+{\cal O}(1\times 10^{-16}),
\qquad
g^{\rm A}_{0\alpha}= {\cal O}(1\times 10^{-21}),
\qquad
g^{\rm A}_{\alpha\beta}= \gamma_{\alpha\beta}+{\cal O}(1\times 10^{-16}).
\label{eq:gab-ACES}
\end{eqnarray}

Furthermore, Eq.~(\ref{eq:x-y_trans_0-ACES}) implies that the proper time of an atomic clock at the origin of the ACRS is nearly equivalent to the time of the SCRS, namely:
\begin{eqnarray}
\frac{d\tau_{\rm A}}{dt_{\rm ISS}}&=&1+{\cal O}(5.6\times 10^{-18}).
\label{eq:y_A_proper-ACES}
\end{eqnarray}

Using the coordinate transformations (\ref{eq:t_ISS})--(\ref{eq:x_ISS}) between the SCRS and the GCRS, we can relate ACES proper time to GCRS geocentric time, TCG. Dropping quadrupole terms from the expression as well as the acceleration-dependent term $c^{-2}{\vec a}_{\rm ISS_0}\cdot {\vec y}_{\rm A0}={\cal O}(4.2\times 10^{-18})$, we obtain
\begin{equation}
\frac{d\tau_{\rm A}}{dt}=
1-\frac{1}{c^2}\Big\{{\textstyle\frac{1}{2}}{\vec v}^2_{\rm A}+
({\vec v}_{\rm A}\cdot{\vec v}_{\rm ad})+
\frac{GM_{\rm E}}{r_{\rm A}}\Big(1+J_2\Big[\frac{R_{\rm E}}{r_{\rm A}}\Big]^2\frac{3z^2_{\rm A}-r^2_{\rm A}}{2r^2_{\rm A}}\Big)
\Big\}+{\cal O}(3\times 10^{-16}).
\label{eq:prop-coord-time-J2*=}
\end{equation}

Clearly, (\ref{eq:prop-coord-time-J2*=}) may be obtained directly by defining the ACRS via a coordinate transformation between the GCRS and ACRS. However, by introducing the hierarchy of reference systems GCRS$\rightarrow$SCRS$\rightarrow$ACRS, we were able to discuss the appropriate gravitational and nongravitational forces and related torques at each step of these transformations.

Note that the accuracy of (\ref{eq:prop-coord-time-J2*=}) may be improved to ${\cal O}(4.4\times 10^{-17})$ by including several terms with the gravitational harmonics beyond $J_2$. Such an improvement may be needed to be consistent with the ultimate ACES clock accuracy, which is expected to be at the level of $1\times 10^{-16}$ at 1 day.

Concluding, we emphasize that, contrary to our original expectations, the ACES coordinate reference system may be treated as a quasi-inertial one. One can use the coordinate transformations and the metric tensor discussed in this section to describe the relativistic observables of the ACES experiment. However,  the proposed Space Optical Clock (SOC) mission \cite{SOC} will require the introduction of a more accurate set of coordinate reference systems.

\section{Modeling the relativistic observables for ACES}
\label{sec:obsere}

Now that we have successfully formulated all the coordinate systems and coordinate transformation rules required to describe the ACES experiment and associated ground stations, we can proceed with modeling the timing and frequency observables of ACES.

\begin{figure}[t]
\begin{minipage}[b]{.49\linewidth}
\includegraphics[width=0.90\linewidth]{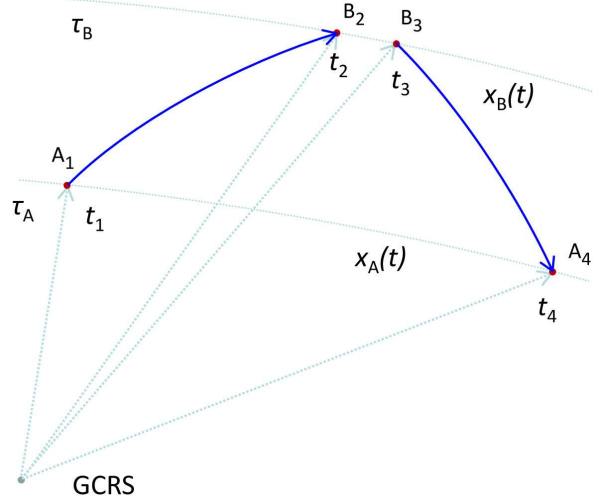}
\end{minipage}
\vskip -2pt
\caption{\label{fig:paired-one-way_events}
Timing events for paired one-way scenarios: Depicted (not to scale) are the trajectories of the terminals $A$ and $B$, with corresponding proper times $\tau_{\rm A}$ and $\tau_{\rm B}$ and with four events in the GCRS, corresponding to a one-way signal transmission at $\vec{x}_{\rm B}(t_1)$ and its reception by the $A$ terminal at $\vec{x}_{\rm A}(t_{2})$, and, similarly, another one-way signal transmission at $\vec{x}_{\rm A}(t_3)$ by terminal $A$ and reception of this signal at $\vec{x}_{\rm B}(t_4)$.
}
\end{figure}

\subsection{Timing observables}
\label{sec:timing_observe}

First, we consider the generic two-way scenario shown in Fig.~\ref{fig:paired-one-way_events}, representative of the European Laser Timing (ELT) experiment \cite{Prochazka-etal:2013}, operating in conjunction with ACES. Depicted are the worldlines of two terminals, $A$ (e.g., a ground station) and $B$ (the ACES experiment). Proper times of transmission and reception, time-stamped by local clocks, are recorded at both locations. Thus, a signal emitted at $\tilde\tau_{\rm A}(t_1)=\tilde\tau_{\rm A1}$ (where the tilde is used to indicate that $\tilde\tau$ is a discrete data point in a measurement sequence, as distinguished from the continuous variable $\tau$ representing proper time) will be recorded at $\tilde\tau_{\rm A1}^{\rm e}=\tilde\tau_{\rm A1}+\delta\tilde\tau_{\rm A}^{\rm e}$, where $\delta\tilde\tau_{\rm A}^{\rm e}$ captures the finite precision of the timestamp and other instrumental uncertainties. We similarly introduce the time of the first signal's reception along worldline $B$, $\tilde\tau_{\rm B2}^{\rm r}=\tilde\tau_{\rm B}(t_2)+\delta\tilde\tau_{\rm B}^{\rm r}=\tilde\tau_{\rm B2}+\delta\tilde\tau_{\rm B}^{\rm r}$, and the corresponding quantities for the return signal: $\tilde\tau_{\rm B3}^{\rm e}=\tilde\tau_{\rm B}(t_3)+\delta\tilde\tau_{\rm B}^{\rm e}=\tilde\tau_{\rm B3}+\delta\tilde\tau_{\rm B}^{\rm e}$ and $\tilde\tau_{\rm A4}^{\rm r}=\tilde\tau_{\rm A}(t_4)+\delta\tilde\tau_{\rm A}^{\rm r}=\tilde\tau_{\rm A4}+\delta\tilde\tau_{\rm A}^{\rm r}$. We assume that $\delta\tilde\tau_{\rm A}^{\rm e}$, $\delta\tilde\tau_{\rm A}^{\rm r}$, $\delta\tilde\tau_{\rm B}^{\rm e}$ and $\delta\tilde\tau_{\rm B}^{\rm r}$ are constant and known ahead of time (through instrumental calibration) though in general, the emission and reception delays are different: $\delta\tilde\tau^{\rm e}_{\rm A}\not=\delta\tilde\tau^{\rm r}_{\rm A}$ and $\delta\tilde\tau^{\rm e}_{\rm B}\not=\delta\tilde\tau^{\rm r}_{\rm B}$. For convenience, we introduce the following combinations corresponding to pseudoranges:
{}
\begin{eqnarray}
\Delta\tilde\tau_{\rm A1B2}&=&\tilde\tau^{\rm r}_{\rm B2}-\tilde\tau^{\rm e}_{\rm A1}
, \qquad
\Delta\tilde\tau_{\rm B3A4}=\tilde\tau^{\rm r}_{\rm A4}-\tilde\tau^{\rm e}_{\rm B3},
\label{eq:timing_comb_range}\\
\Delta\tilde\tau_{\rm A1A4}&=&\tilde\tau^{\rm r}_{\rm A4}-\tilde\tau^{\rm e}_{\rm A1}, \qquad
\Delta\tilde\tau_{\rm B2B3}=\tilde\tau^{\rm e}_{\rm B3}-\tilde\tau^{\rm r}_{\rm B2}.
\label{eq:timing_comb_delay}
\end{eqnarray}

These values are all functions of the coordinate time quadruplet of coordinate times $\{t_1,t_2,t_3,t_4\}$. Using Eq.~(\ref{eq:light-time}), we can use these coordinate times to form the light travel times
\begin{eqnarray}
\Delta t_{12}=t_2-t_1&=&c^{-1}{\cal R}_{\rm AB}\big({\vec x}_{\rm B1},{\vec x}_{\rm A2}\big), \qquad
\Delta t_{34}=t_4-t_3=c^{-1}{\cal R}_{\rm AB}\big({\vec x}_{\rm B3},{\vec x}_{\rm A4}\big),
\label{eq:synchron}
\end{eqnarray}
where
\begin{eqnarray}
{\cal R}_{\rm AB}\big({\vec x}_{\rm A1},{\vec x}_{\rm B2}\big)&=& |{\vec x}_{\rm B2} - {\vec x}_{\rm A1}|+(1+\gamma)\frac{G M_{\rm E}}{c^2}
\ln\left[\frac{r_{\rm A1}+r_{\rm B2}+|{\vec x}_{\rm B2} - {\vec x}_{\rm A1}|}{r_{\rm A1}+r_{\rm B2}-|{\vec x}_{\rm B2} - {\vec x}_{\rm A1}|}\right]+{\cal O}\big(c^{-3}, J_2, G^2\big),
\label{eq:R-total0}\\
{\cal R}_{\rm AB}\big({\vec x}_{\rm B3},{\vec x}_{\rm A4}\big)&=& |{\vec x}_{\rm A4} - {\vec x}_{\rm B3}|+(1+\gamma)\frac{G M_{\rm E}}{c^2}
\ln\left[\frac{r_{\rm A4}+r_{\rm B3}+|{\vec x}_{\rm A4} - {\vec x}_{\rm B3}|}{r_{\rm A4}+r_{\rm B3}-|{\vec x}_{\rm A4} - {\vec x}_{\rm B3}|}\right]+{\cal O}\big(c^{-3}, J_2, G^2\big),
\label{eq:R-total1}
\end{eqnarray}
where we used the shorthand $\vec{x}_{\rm A1}=\vec{x}_{\rm A}(t_{1})$, $\vec{x}_{\rm B2}=\vec{x}_{\rm B}(t_{\rm 2})$, $\vec{x}_{\rm B3}=\vec{x}_{\rm B}(t_{\rm 3})$, and $\vec{x}_{\rm A4}=\vec{x}_{\rm A}(t_{4})$ to indicate the various positions along the two worldlines. Additionally, we can form the time intervals $\Delta t_{14}=t_4-t_1$ and $\Delta t_{23}=t_3-t_2$, which correspond to the intervals of proper time measured by the clocks on $A$ and $B$.

\subsubsection{Clock synchronisation}
The quantities (\ref{eq:timing_comb_range})--(\ref{eq:timing_comb_delay}) together with
(\ref{eq:synchron}) and (\ref{eq:R-total0})--(\ref{eq:R-total1}) allow us to express $\Delta t=t_3-t_1$, which is required for synchronization of two clocks \cite{Blanchet-etal:2001,Turyshev:2012nw,ref:Turyshev:2014dea}:
\begin{eqnarray}
\Delta t=t_3-t_1&=&\frac{1}{2}\Big(\Delta t_{14}+\Delta t_{23}+\Delta t_{12}-\Delta t_{34}\Big).
\label{eq:synch}
\end{eqnarray}
$\Delta t$ is known because it is expressed in terms of $\Delta t_{14}$  and $\Delta t_{23}$, which are measured at the two terminals, and $\Delta t_{12}$ and $\Delta t_{34}$, which are computed from Eqs.~(\ref{eq:synchron})--(\ref{eq:R-total1}).

The sum of the two Eqs.~(\ref{eq:synchron}) may also be written as
\begin{eqnarray}
t_3-t_1&=& \frac{1}{2}\Big((t_4-t_1)+(t_3-t_2)+c^{-1}{\cal R}_{\rm AB}\big({\vec x}_{\rm B1},{\vec x}_{\rm A2}\big)-c^{-1}{\cal R}_{\rm AB}\big({\vec x}_{\rm A3},{\vec x}_{\rm B4}\big)\Big).
\label{eq:synchron2}
\end{eqnarray}
The difference between the two range measurements ${\cal R}_{\rm AB}({\vec x}_{\rm B1},{\vec x}_{\rm A2})$ and ${\cal R}_{\rm AB}({\vec x}_{\rm A3},{\vec x}_{\rm B4})$ produces a correction to the the sum of the two clock times intervals measured at the two terminals. This correction depends on the range-rate between the two terminals \cite{Degnan-2002}. Equation (\ref{eq:synchron2}) can also be written as
\begin{align}
\frac{1}{2}(t_1+t_4)-\frac{1}{2}(t_2+t_3)=\frac{1}{2c}\Big({\cal R}_{\rm AB}\big({\vec x}_{\rm A3},{\vec x}_{\rm B4}\big)-{\cal R}_{\rm AB}\big({\vec x}_{\rm B1},{\vec x}_{\rm A2}\big)\Big),\label{eq:synchron3}
\end{align}
which represents the essence of Einstein's procedure for clock synchronization.

Clearly, for ideal clocks and perfect measurements, (\ref{eq:synchron3}) is an exact identity. However, for realistic measurements, when various sources of noise  present in the system, (\ref{eq:synchron3}) could lead to an observational model that may be used to evaluate the stability of clock synchronization \cite{Abshire-Gardner:1985,Samain-etal:2015,Exertier-etal:2014}. The baseline approach in ACES is to use the MWL hardware to enable time and frequency transfer \cite{Delva-etal:2012}. Alternatively, the ELT may also be used for this purpose, relying only on the timing information in Eqs.~(\ref{eq:timing_comb_range})--(\ref{eq:timing_comb_delay}). Below, we develop the appropriate relativistic models.

To express the observed clock offset, we use the quadruplet $\{\tilde\tau^{\rm e}_{\rm A1}, \tilde\tau^{\rm r}_{\rm B2}, \tilde\tau^{\rm e}_{\rm B3}, \tilde\tau^{\rm r}_{\rm A4}\}$ of the time tags reported by terminals $A$ and $B$:
\begin{eqnarray}
\delta\tau^{\rm obs}_{\rm AB}&=&\frac{1}{2}\Big((\tilde\tau^{\rm e}_{\rm A1}+\tilde\tau^{\rm r}_{\rm A4})-(\tilde\tau^{\rm r}_{\rm B2}+\tilde\tau^{\rm e}_{\rm B3})\Big).
\label{eq:de-offest-obs}
\end{eqnarray}
This term can also be computed as follows:
{}
\begin{eqnarray}
\delta\tau^{\rm comp}_{\rm AB}&=&\frac{1}{2}\Big((\hat\tau^{\rm }_{\rm A1}-\tau^{\rm }_{\rm A1})+(\tau^{\rm }_{\rm A1}-t_1)+(\hat\tau^{\rm }_{\rm A4}-\tau^{\rm }_{\rm A4})+(\tau^{\rm }_{\rm A4}-t_4)+\delta\tau^{\rm e}_{\rm A}+\delta\tau^{\rm r}_{\rm A}\Big)-\nonumber\\
&-&\frac{1}{2}\Big((\hat\tau^{\rm }_{\rm B2}-\tau^{\rm }_{\rm B2})+(\tau^{\rm }_{\rm B2}-t_2)+(\hat\tau^{\rm }_{\rm B3}-\tau^{\rm }_{\rm B3})+(\tau^{\rm }_{\rm B3}-t_3)+\delta\tau^{\rm e}_{\rm B}+\delta\tau^{\rm r}_{\rm B}\Big)+\nonumber\\
&+&\frac{1}{2}\Big((t_4-t_3)-(t_2-t_1)\Big)+\frac{1}{2c}\big(\delta^{\rm atm}_{34}-\delta^{\rm atm}_{12}\big),
\label{eq:de-offest-comp2}
\end{eqnarray}
where the terms in the equation above represent five groups of terms that have the following meaning:
\begin{itemize}
  \item The quantities $(\hat\tau^{\rm }_{\rm A1}-\tau^{\rm }_{\rm A1}), (\hat\tau^{\rm }_{\rm A4}-\tau^{\rm }_{\rm A4})$ and $(\hat\tau^{\rm }_{\rm B2}-\tau^{\rm }_{\rm B2}), (\hat\tau^{\rm }_{\rm B3}-\tau^{\rm }_{\rm B3})$ represent the differences between recorded time tags and proper times. These clock errors may be modeled as quadratic functions of the measured proper time. For example, the time tag error $(\tilde\tau^{\rm }_{\rm A1}-\tau^{\rm }_{\rm A1})$ is given as:
\begin{equation}
(\hat\tau^{\rm }_{\rm A1}-\tau^{\rm }_{\rm A1})=\alpha_{\rm A}+\beta_{\rm A}(\hat\tau^{\rm }_{\rm A1}-\hat\tau^{\rm }_{\rm A0})+\textstyle{\frac{1}{2}}\gamma_{\rm A}(\hat\tau^{\rm }_{\rm A1}-\hat\tau^{\rm }_{\rm A0})^2+{\cal O}(\Delta\hat\tau^{\rm 3}_{\rm A1}),
\label{eq:tau-hat}
\end{equation}
 where $\hat\tau^{\rm }_{\rm A0}$ is a specified epoch and $\Delta\hat\tau^{\rm }_{\rm A1}=\hat\tau^{\rm }_{\rm A1}-\hat\tau^{\rm }_{\rm A0}$. The constants $\alpha_{\rm A}, \beta_{\rm A}$ and $\gamma_{\rm A}$ are to be estimated during the data analysis. The other time tag errors are modeled in a similar manner.
  \item The quantities $(\tau^{\rm }_{\rm A1}-t_1), (\tau^{\rm }_{\rm A4}-t_4)$ and $(\tau^{\rm }_{\rm B2}-t_2), (\tau^{\rm }_{\rm B3}-t_3)$ are the general relativistic differences between proper time and coordinate time. These relationships are established by integrating the equations (\ref{eq:proper-coord-t+}) and (\ref{eq:prop-coord-time-J2*=}).
  \item The error terms $\delta\tau^{\rm e}_{\rm A}, \delta\tau^{\rm r}_{\rm A}$ and $\delta\tau^{\rm e}_{\rm B}, \delta\tau^{\rm r}_{\rm B}$ are instrumental delays due to fixed propagation paths between the optical systems, detector electronics and the time tag unit. These delays are usually measured and  calibrated before the flight.
  \item The terms $(t_4-t_3)$ and $(t_2-t_1)$ represent the coordinate light transfer time elapsed between the signal emission at one terminal and its reception at the opposing terminal, related to pseudo ranges as given by (\ref{eq:synchron}).
  \item The term $(\delta^{\rm atm}_{34}-\delta^{\rm atm}_{12})$ is the difference the atmospheric signal propagation delay during uplink and downlink.
\end{itemize}

In the case of the ELT experiment, the signal received on the ISS is time-stamped at the reception and is simultaneously returned by a retroreflector. Thus, $\tilde\tau^{\rm r}_{\rm B2}=\tilde\tau^{\rm e}_{\rm B3}$, $\Delta\tilde\tau_{\rm B2B3}=0$ and (\ref{eq:de-offest-obs}) becomes:
{}
\begin{eqnarray}
\delta\tau^{\rm obs}_{\rm AB}&=&\textstyle{\frac{1}{2}}(\tilde\tau^{\rm e}_{\rm A1}+\tilde\tau^{\rm r}_{\rm A4})-\tilde\tau^{\rm r}_{\rm B2}.
\label{eq:de-offest-obs_ACES}
\end{eqnarray}
Furthermore, $\tilde\tau^{\rm r}_{\rm B2}=\tilde\tau^{\rm e}_{\rm B3}$, and, therefore, (\ref{eq:de-offest-comp2}) becomes:
{}
\begin{eqnarray}
\delta\tau^{\rm comp}_{\rm AB}&=&\frac{1}{2}\Big((\hat\tau^{\rm }_{\rm A1}-\tau^{\rm }_{\rm A1})+(\tau^{\rm }_{\rm A1}-t_1)+(\hat\tau^{\rm }_{\rm A4}-\tau^{\rm }_{\rm A4})+(\tau^{\rm }_{\rm A4}-t_4)+\delta\tau^{\rm e}_{\rm A}+\delta\tau^{\rm r}_{\rm A}\Big)-
\nonumber\\
&&-~
\Big((\hat\tau^{\rm }_{\rm B2}-\tau^{\rm }_{\rm B2})+(\tau^{\rm }_{\rm B2}-t_2)+\delta\tau^{\rm r}_{\rm B}\Big)+\nonumber\\
&&+~
\frac{1}{2c}\Big({\cal R}_{\rm AB}\big({\vec x}_{\rm B3},{\vec x}_{\rm A4}\big)-{\cal R}_{\rm AB}\big({\vec x}_{\rm A1},{\vec x}_{\rm B2}\big)\Big)+\frac{1}{2c}\big(\delta^{\rm atm}_{34}-\delta^{\rm atm}_{12}\big).
\label{eq:de-offest-comp2_ACES}
\end{eqnarray}
This model could be used to evaluate clock desynchronization in the ACES experiment. It may also be  used to develop appropriate simulation code that is needed to understand the features of ACES. Note that when two ground-based clocks are used for common view comparison (see Table~\ref{tb:caps}) in a terrestrial reference frame, the Sagnac effect must also be accounted for, as discussed in \cite{Ashby-2003}.

The first $1/c$-term in (\ref{eq:de-offest-comp2_ACES}) represents the difference between optical paths (given by (\ref{eq:R-total0}) and (\ref{eq:R-total1})) of the two signals traveling between receiver and transmitter in a two-way communication link.  This term depends on the geocentric positions of both the ACES and a ground station. This information, in particular, is helpful to investigate the accuracy needed for trajectory reconstruction of the ISS in order to satisfy the ACES' science requirements. Similar questions were addressed previously under different sets of assumptions \cite{Duchayne-etal:2009,Schreiber-etal:2009,Montenbruck-etal:2011,Wermuth-etal:2012}.

Using (\ref{eq:de-offest-comp2_ACES}), we evaluate the geometric uncertainty due to imprecision in the ISS navigation:
{}
\begin{eqnarray}
\delta\tau^{\rm comp}_{\rm AB}\big|_{\rm geom}&=&
\frac{1}{2c}\Big({\cal R}_{\rm AB}\big({\vec x}_{\rm B3},{\vec x}_{\rm A4}\big)-{\cal R}_{\rm AB}\big({\vec x}_{\rm A1},{\vec x}_{\rm B2}\big)\Big)=\frac{1}{2c}\Big(|{\vec x}_{\rm A4} - {\vec x}_{\rm B3}|-|{\vec x}_{\rm B2} - {\vec x}_{\rm A1}|\Big),
\label{eq:de-offest-geom-ACES}
\end{eqnarray}
where we neglected the contribution from the Shapiro delay, which, even in the absolute sense, contributes $\leq10$~ps. For the ELT experiment, where $t_3=t_2$, (\ref{eq:de-offest-geom-ACES}) becomes
{}
\begin{eqnarray}
\delta\tau^{\rm comp}_{\rm AB}\big|_{\rm geom}&=&
\frac{1}{2c}({\vec n}_{\rm AB}\cdot{\vec v}_{\rm AB})(t_4-t_1)+{\cal O}(c^{-2}).
\label{eq:de-offest-geom-ACES*}
\end{eqnarray}
Consider the case in which the velocity of the ground-based station is well known and the largest systematic error comes from the uncertainty in the velocity of the ISS. We take the instant with the longest round-trip time $\Delta t_{14}=t_4-t_1$ of $\Delta t_{14}^{\rm max}=2c^{-1}\sqrt{(r_{\rm E}+h_{\rm ISS})^2 -r_{\rm E}^2}\approx 2c^{-1}\sqrt{2r_{\rm E}h_{\rm ISS}}=15.2$~ms. The angle between two unit vectors $\hat{\vec n}_{\rm AB}$ and $\hat{\vec v}_{\rm B}$ at that instant is also small, $\cos(\hat{\vec n}_{\rm AB}, \hat{\vec v}_{\rm B}) \approx 1-h_{\rm ISS}/r_{\rm E}$. Then, assuming that the timing accuracy for the ISS pass is better than $\delta\tau^{\rm comp}_{\rm AB}\big|_{\rm geom}\leq0.3$~ps, from (\ref{eq:de-offest-geom-ACES*}) we have the accuracy at which the ISS velocity needs to be known:
{}
\begin{eqnarray}
\Delta v_{\rm B}&\leq& \frac{c^2}{\sqrt{2r_{\rm E}h_{\rm ISS}}}0.3~{\rm ps}=0.012 ~{\rm m/s},
\label{eq:v-req-ACES}
\end{eqnarray}
which is equivalent to the requirement that the geocentric position of the ACES package on the ISS be known to $\Delta h_{\rm ISS}\leq2(r_{\rm E}+h_{\rm ISS})\Delta v_{\rm B}/v_{\rm ISS}=20.8$~m, assuming that this positional error is responsible for the entire timing error of 0.3~ps. On the other hand, as we see from (\ref{eq:de-offest-comp2_ACES}), there will be many contributions to the timing error, with the ISS positional error being just one of them. Thus, based on the anticipated precision of the time transfer experiments, we require that the ISS positional error contributes less than 10\% of the timing error of 0.3~ps, which translates into a requirement that the geocentric position of ACES be known to $\sigma_{h_{\rm ISS}}\leq\sqrt{0.1}\cdot20.8~{\rm m}=6.3$~m. This is consistent with earlier estimates \cite{Duchayne-etal:2009} and can be achieved with existing navigation capabilities \cite{Montenbruck-etal:2011}.

\subsubsection{Two-way pseudorange}

The time observables (\ref{eq:timing_comb_range})--(\ref{eq:timing_comb_delay}) may also be used to develop a two-way pseudorange estimate. Similarly to (\ref{eq:de-offest-obs}), we use the time tag quadruplet $\{\tilde\tau^{\rm e}_{\rm A1}, \tilde\tau^{\rm r}_{\rm B2}, \tilde\tau^{\rm e}_{\rm B3}, \tilde\tau^{\rm r}_{\rm A4}\}$ to derive an expression for the observed two-way pseudorange, which is half the sum of the one-way pseudoranges:
{}
\begin{eqnarray}
\delta\rho^{\rm obs}_{\rm AB}&=&\textstyle{\frac{1}{2}}c\Big(\Delta\tilde\tau_{\rm A1B2}+\Delta\tilde\tau_{\rm B3A4}\Big)=\textstyle{\frac{1}{2}}c\Big(\Delta\tilde\tau_{\rm A1A4}-\Delta\tilde\tau_{\rm B2B3}\Big)\equiv\frac{1}{2}c\Big((\tilde\tau^{\rm e}_{\rm A4}-\tilde\tau^{\rm r}_{\rm A1})-(\tilde\tau^{\rm r}_{\rm B3}-\tilde\tau^{\rm e}_{\rm B2})\Big),
\label{eq:range-obs}
\end{eqnarray}
where all the times are the actual time measurements reported by the two terminals. In the case of the ELT experiment \cite{Prochazka-etal:2013}, $\tilde\tau^{\rm r}_{\rm B2}=\tilde\tau^{\rm e}_{\rm B3}$,
and Eq.~(\ref{eq:range-obs}) becomes:
{}
\begin{eqnarray}
\delta\tau^{\rm obs}_{\rm AB}&=&\textstyle{\frac{1}{2}}c(\tilde\tau^{\rm e}_{\rm A4}-\tilde\tau^{\rm r}_{\rm A1}).
\label{eq:range-obs_ACES}
\end{eqnarray}

The computed pseudorange is given by
{}
\begin{eqnarray}
\delta\rho^{\rm comp}_{\rm AB}&=&\textstyle{\frac{1}{2}}c\Big(\Delta\hat\tau_{\rm A1B2}+\Delta\hat\tau_{\rm B3A4}\Big)\equiv\frac{1}{2}c\Big((\hat\tau^{\rm r}_{\rm B2}-\hat\tau^{\rm e}_{\rm A1})+(\hat\tau^{\rm r}_{\rm A4}-\hat\tau^{\rm e}_{\rm B3})\Big),
\label{eq:range-comp}
\end{eqnarray}
where now times are modeled times for those reported by the clocks on stations $A$ and $B$. To facilitate computation, this expression is written as a sum of the following terms:
{}
\begin{eqnarray}
\delta\rho^{\rm comp}_{\rm AB}&=&\textstyle{\frac{1}{2}}c\Big((\hat\tau^{\rm }_{\rm A4}-\tau^{\rm }_{\rm A4})+(\tau^{\rm }_{\rm A4}-t_4)-(\hat\tau^{\rm }_{\rm A1}-\tau^{\rm }_{\rm A1})-(\tau^{\rm }_{\rm A1}-t_1)+\delta\tau^{\rm r}_{\rm A}-\delta\tau^{\rm e}_{\rm A}\Big)-\nonumber\\
&-&\textstyle{\frac{1}{2}}c\Big((\hat\tau^{\rm }_{\rm B3}-\tau^{\rm }_{\rm B3})+(\tau^{\rm }_{\rm B3}-t_3)-(\hat\tau^{\rm }_{\rm B2}-\tau^{\rm }_{\rm B2})-(\tau^{\rm }_{\rm B2}-t_2)+\delta\tau^{\rm e}_{\rm B}-\delta\tau^{\rm r}_{\rm B}\Big)+\nonumber\\
&+&\textstyle{\frac{1}{2}}c\Big((t_4-t_3)+(t_2-t_1)\Big)+\frac{1}{2}\big(\delta^{\rm atm}_{12}+\delta^{\rm atm}_{34}\big),
\label{eq:range-comp2}
\end{eqnarray}
where all the quantities present in this equation are defined after (\ref{eq:de-offest-comp2}).

For the ELT experiment, $\tilde\tau^{\rm r}_{\rm B2}=\tilde\tau^{\rm e}_{\rm B3}$, and, therefore, (\ref{eq:range-comp2}) becomes:
{}
\begin{eqnarray}
\delta\rho^{\rm comp}_{\rm AB}&=&\textstyle{\frac{1}{2}}c\Big((\hat\tau^{\rm }_{\rm A4}-\tau^{\rm }_{\rm A4})+(\tau^{\rm }_{\rm A4}-t_4)-(\hat\tau^{\rm }_{\rm A1}-\tau^{\rm }_{\rm A1})-(\tau^{\rm }_{\rm A1}-t_1)+\delta\tau^{\rm r}_{\rm A}-\delta\tau^{\rm e}_{\rm A}\Big)-\nonumber\\
&+&\textstyle{\frac{1}{2}}\Big({\cal R}_{\rm AB}\big({\vec x}_{\rm A1},{\vec x}_{\rm B2}\big)+{\cal R}_{\rm AB}\big({\vec x}_{\rm B2},{\vec x}_{\rm A4}\big)\Big)+\frac{1}{2}\big(\delta^{\rm atm}_{12}+\delta^{\rm atm}_{24}\big).
\label{eq:range-comp2-ACES}
\end{eqnarray}

The two-way range model (\ref{eq:range-comp2-ACES}) may be used to improve navigation of the ACES package on the ISS. If enough satellite laser ranging (SLR) stations participate in the laser ranging campaign, ACES may be able to completely address its precision navigation needs. Although geodetic satellites yield trajectory reconstruction accuracy at the level of $\sim$1~cm, the ISS will not allow for such a precision. As we discussed earlier, the station is subject to significant nongravitational forces whose presence degrades the trajectory reconstruction. Nevertheless, because of its potential navigational value, the use of SLR for ACES needs further investigation, especially within the ELT experiment \cite{Schreiber-etal:2009}.

\subsection{Frequency observables: the gravitational redshift}
\label{sec:freq-obs}

The microwave link (MWL) component of ACES utilizes three different microwave frequencies (one uplink, two downlink) to provide for reliable (not weather-dependent) transfer of timing and frequency information. In particular, for frequency observables, the combined use of uplink and downlink observables allows for the formulation of a science observable, from which the effects of the Doppler frequency shift and atmospheric noise are canceled to the first order.

To formulate the gravitational redshift observable, we assume the presence of a one-way and a two-way frequency observable. Two-way, in this context, means a coherent retransmission of a signal received from the ground. We assume that the signal retransmission is instantaneous. Such an observable can always be synthesized using the three radio frequency observables that are produced by the MWL experiment.

We use $\hat{\tau}_{\rm A1}$ to represent the proper time transmission of a signal from worldline $A$ at proper time $\hat\tau_{\rm A}(t_1)$, which is then received at worldline $B$ at proper time $\hat{\tau}_{\rm B2}$. An infinitesimal interval $d\hat{\tau}_{\rm A1}$ at the transmitter corresponds to an infinitesimal interval $d\hat{\tau}_{\rm B2}$ at the receiver. The number of cycles transmitted during this interval, $dn$, is equal to the number of cycles received at the receiver. On the other hand the number of cycles is the product of the signal frequency and the time interval, therefore $f_{\rm A}^{\rm tx}(\hat\tau_{\rm A1})d\hat\tau_{\rm A1}=f_{\rm B}^{\rm rx}(\hat\tau_{\rm B2})d\hat\tau_{\rm B2}$, or
\begin{align}
f_{\rm B}^{\rm rx}(\hat\tau_{\rm B2})=f_{\rm A}^{\rm tx}(\hat\tau_{\rm A1})\frac{d\hat\tau_{\rm A1}}{d\hat\tau_{\rm B2}}.
\end{align}
Similarly, this signal is immediately retransmitted to the ground at the same frequency and then received at $\hat\tau_{\rm A3}$ in a two-way retransmission scheme, the received frequency is given by
\begin{align}
f_{\rm A}^{\rm 2w}(\hat\tau_{\rm A3})=f_{\rm B}^{\rm rx}(\hat\tau_{\rm B2})\frac{d\hat\tau_{\rm B2}}{d\hat\tau_{\rm A3}}=
f_{\rm A}^{\rm tx}(\hat\tau_{\rm A1})\frac{d\hat\tau_{\rm A1}}{d\hat\tau_{\rm B2}}\frac{d\hat\tau_{\rm B2}}{d\hat\tau_{\rm A3}}.
\end{align}
Similarly, a one-way signal transmitted from $B$ at $\hat\tau_{\rm B2}$ at frequency $f_{\rm B}^{\rm tx}(\hat\tau_{\rm B2})$ is received at station $A$ at the frequency
\begin{align}
f_{\rm A}^{\rm 1w}(\hat\tau_{\rm A3})=f_{\rm B}^{\rm tx}(\hat\tau_{\rm B2})\frac{d\hat\tau_{\rm B2}}{d\hat\tau_{\rm A3}}.
\end{align}
The one-way frequency shift between the transmission and the reception frequency includes contributions from the relativistic Doppler effect, from atmospheric noise, and from the gravitational redshift. In contrast, the frequency shift in the two-way transmission includes twice the contributions from the relativistic Doppler effect and from atmospheric noise, but gravitational redshift contributions are canceled out to first order \cite{Blanchet-etal:2001,Vessot1980}. This allows for the formulation of a first-order ``Doppler-canceled'' observable in the form,
\begin{eqnarray}
\delta \eta^{\rm obs}_f=
\frac{\hat f^{\rm 1w}_{\rm B}(\hat\tau_{\rm A3})- \hat f_{\rm B}(\hat\tau_{\rm B2})}{\hat f_{\rm B}(\hat\tau_{\rm B2})}-
\frac{1}{2}\frac{\hat f^{\rm 2w}_{\rm A}(\hat\tau_{\rm A3})-\hat f_{\rm A}(\hat\tau_{\rm A1})}{ \hat f_{\rm A}(\hat\tau_{\rm A1})}.
\label{eq:Doppler-can_obs=}
\end{eqnarray}
For ACES, the removal of the first-order Doppler effect will be enabled by the MWL \cite{Delva-etal:2012}, which will also allow for the removal of the troposphere time delay, as well as the removal of instrumental delays and common mode effects. 

This observable can also be computed from the ratio of proper times that, in turn, can be estimated from orbits:
\begin{eqnarray}
\delta\eta^{\rm comp}_f&=&
\Big[\frac{d\hat\tau_{\rm B2}}{d\hat\tau_{\rm A3}}-1\Big]-\frac{1}{2}\Big[\frac{d\hat\tau_{\rm A1}}{d\hat\tau_{\rm A3}}-1\Big].
\label{eq:Doppler-can_comp*}
\end{eqnarray}

To develop Eq.~(\ref{eq:Doppler-can_comp*}) further, we use the differential equation that relates the rate of the proper times, $\tau_{\rm A}$ and $\tau_{\rm B}$, as measured by a ground-based and on-board clock in Earth's orbit, correspondingly, to the time in GCRS, denoted here as $t$ (see Ref.~\cite{Turyshev-Toth:2013}) as
{}
\begin{eqnarray}
\frac{d\tau_{\rm A}}{dt}&=& 1-\frac{1}{c^2}\Big[\frac{{\vec v}^2_{\rm A}}{2}+U_{\rm E}({\vec y}_{\rm A})\Big]+{\cal O}(c^{-4}) ~~~~{\rm and}~~~~
\frac{d\tau_{\rm B}}{dt}= 1-\frac{1}{c^2}\Big[\frac{{\vec v}^2_{\rm B}}{2}+U_{\rm E}({\vec y}_{\rm B})\Big]+{\cal O}(c^{-4}).
\label{eq:proper-coord-t}
\end{eqnarray}
Taking into account that $({d\tau^{\rm }_{\rm A}}/{dt}-1)\approx ({d\tau^{\rm }_{\rm B}}/{dt}-1)\sim 10^{-9}$, and assuming that time tag errors, $\hat\tau_{\rm A}$, and instrumental drifts, $\delta\tau^{\rm e}_{\rm A}$, are small, we can rewrite $d\hat\tau_{\rm A1}$ in the following form:
{}
\begin{eqnarray}
d\hat\tau_{\rm A1}&=&\Big(1+\frac{d(\hat\tau_{\rm A}-\tau^{\rm }_{\rm A})}{dt}+\frac{d\delta\tau^{\rm e}_{\rm A}}{dt}\Big)\Big(\frac{d\tau^{\rm }_{\rm A}}{dt}\Big)_{t_1}dt_1+{\cal O}(\epsilon_{\rm A}),
\label{eq:time-tags_comp1}
\end{eqnarray}
were we use $({d\tau_{\rm A}}/{dt})_{t_1}$ to mean the value of the expression (\ref{eq:proper-coord-t}) at $t=t_1$ and the error term, $\epsilon_{\rm A}$, is
{}
\begin{eqnarray}
\epsilon_{\rm A}=-\Big[\frac{d\tau^{\rm }_{\rm A}}{dt}-1\Big]\Big(\frac{d(\hat\tau_{\rm A}-\tau^{\rm }_{\rm A})}{dt}+\frac{d\delta\tau^{\rm e}_{\rm A}}{dt}\Big)+{\cal O}(c^{-4}).
\label{eq:time-tags_comp2}
\end{eqnarray}
For $\epsilon_{\rm A}$ to be less than the frequency stabilization accuracy anticipated from on ACES, i.e., $\epsilon_{\rm A}\leq 10^{-17}$, the time tag errors (i.e., scale and clock acceleration errors), $\delta\dot{\hat\tau}_{\rm A}={d(\hat\tau_{\rm A}-\tau^{\rm }_{\rm A})}/{dt}$,  and instrumental drifts, $\delta\dot\tau^{\rm e}_{\rm A}={d\delta\tau^{\rm e}_{\rm A}}/{dt}$, must not exceed $\delta\dot{\hat\tau}_{\rm A}, \delta\dot\tau^{\rm e}_{\rm A}\leq 10$~ns/s. Otherwise the approximation needs to be updated to include (\ref{eq:time-tags_comp2}), which we omit below. Using (\ref{eq:tau-hat}), our approximation results in
{}
\begin{equation}
\delta\dot{\hat\tau}_{\rm A}\equiv\frac{d(\hat\tau_{\rm A}-\tau^{\rm }_{\rm A})}{dt}=\beta_{\rm A}+\gamma_{\rm A}(\hat\tau^{\rm }_{\rm A}-\hat\tau^{\rm }_{\rm A0})+{\cal O}(\Delta\hat\tau^{\rm 2}_{\rm A}, c^{-2}).
\label{eq:tau-hat-dot}
\end{equation}
As a result, to approximation appropriate for ACES, (\ref{eq:time-tags_comp1}) takes the following linearized form:
{}
\begin{eqnarray}
d\hat\tau_{\rm A1}&=&\Big(1+\delta\dot{\hat\tau}_{\rm A}+\delta\dot\tau^{\rm e}_{\rm A}\Big)\Big(\frac{d\tau^{\rm }_{\rm A}}{dt}\Big)_{t_1}dt_1.
\label{eq:time-tags_comp4}
\end{eqnarray}
Similarly, we have the other two expressions
\begin{eqnarray}
d\hat\tau_{\rm A3}&=&\Big(1+\delta\dot{\hat\tau}_{\rm A}+\delta\dot\tau^{\rm r}_{\rm A}\Big)\Big(\frac{d\tau^{\rm }_{\rm A}}{dt}\Big)_{t_3}dt_1,
\\
d\hat\tau_{\rm B2}&=&\Big(1+\delta\dot{\hat\tau}_{\rm B}+\delta\dot\tau^{\rm r}_{\rm B}\Big)\Big(\frac{d\tau^{\rm }_{\rm B}}{dt}\Big)_{t_2}dt_1.
\label{eq:eq:proper-coord-t_A}
\end{eqnarray}

Therefore, the ratio of the estimates of proper times in (\ref{eq:Doppler-can_comp*}) may be expressed via the ratio of their coordinate counterparts as
{}
\begin{eqnarray}
\frac{d\hat\tau_{\rm B2}}{d\hat\tau_{\rm A3}}
&=&\Big(1+\delta\dot{\hat\tau}_{\rm B}+\delta\dot\tau^{\rm r}_{\rm B}-\delta\dot{\hat\tau}_{\rm A}-\delta\dot\tau^{\rm r}_{\rm A}\Big)\Big(\frac{d\tau_{\rm B}}{dt}\Big)_{t_2}\Big(\frac{d\tau_{\rm A}}{dt}\Big)^{-1}_{t_3}\frac{dt_{\rm 2}}{dt_{\rm 3}},
\\
\frac{d\hat\tau_{\rm A1}}{d\hat\tau_{\rm A3}}
&=&\Big(1+\delta\dot\tau^{\rm e}_{\rm A}-\delta\dot\tau^{\rm r}_{\rm A}\Big)\Big(\frac{d\tau_{\rm A}}{dt}\Big)_{t_1}\Big(\frac{d\tau_{\rm A}}{dt}\Big)^{-1}_{t_3}\frac{dt_{\rm 1}}{dt_{\rm 3}}.
\label{eq:tauAB}
\end{eqnarray}

As a result, Eq.~(\ref{eq:Doppler-can_comp*}) takes the form
{}
\begin{eqnarray}
\delta\eta^{\rm comp}_f=
\Big(\frac{dt_{\rm 2}}{dt_{\rm 3}}-1\Big)-\frac{1}{2}\Big(\frac{dt_{\rm 1}}{dt_{\rm 3}}-1\Big)&+&
\Big[\Big(1+\delta\dot{\hat\tau}_{\rm B}+\delta\dot\tau^{\rm r}_{\rm B}-\delta\dot{\hat\tau}_{\rm A}-\delta\dot\tau^{\rm r}_{\rm A}\Big)\Big(\frac{d\tau_{\rm B}}{dt}\Big)_{t_2}\Big(\frac{d\tau_{\rm A}}{dt}\Big)^{-1}_{t_3}-1\Big]\frac{dt_{\rm 2}}{dt_{\rm 3}}-
\nonumber\\
&-&
\frac{1}{2}\Big[\Big(1+\delta\dot\tau^{\rm e}_{\rm A}-\delta\dot\tau^{\rm r}_{\rm A}\Big)\Big(\frac{d\tau_{\rm A}}{dt}\Big)_{t_1}\Big(\frac{d\tau_{\rm A}}{dt}\Big)^{-1}_{t_3}-1\Big]\frac{dt_{\rm 1}}{dt_{\rm 3}},
\label{eq:Doppler-canceled}
\end{eqnarray}
where the instances of coordinate time $t_2$ and $t_3$ are related by (\ref{eq:synchron}). Although events of the original signal emission at $t_{\rm 1}$ and its ultimate reception at $t_{\rm 3}$ are not connected by the same light cone, we nevertheless may compute the total time elapsed between the two events (as was first observed in \cite{Fock-book:1959}). Indeed, taking $t_2=t_3$ in (\ref{eq:synchron}), we have:
{}
\begin{equation}
t_{\rm 1}=t_{\rm 3} - c^{-1}\Big({\cal R}_{\rm AB}\big({\vec x}_{\rm A}(t_1),{\vec x}_{\rm B}(t_2)\big)+{\cal R}_{\rm BA}\big({\vec x}_{\rm B}(t_2),{\vec x}_{\rm A}(t_3)\big)\Big).
\label{eq:transm-time:ABA}
\end{equation}
Thus, the total coordinate time elapsed between the two events is fully determined by the locations of the two stations. In fact, using  the first equation in (\ref{eq:synchron}) and (\ref{eq:transm-time:ABA}) we have the following exact expression for the ratio of coordinate times present in (\ref{eq:Doppler-canceled}):
{}
\begin{eqnarray}
\frac{dt_{\rm 2}}{dt_{\rm 3}}&=&1-\frac{1}{c}\frac{d}{dt_{\rm 3}}\Big[{\cal R}_{\rm BA}({\vec x}_{\rm B2},{\vec x}_{\rm A3})\Big]-\frac{1}{c}\frac{d\delta^{\rm atm}_{\rm B23}}{dt_{\rm 3}},
\label{eq:dt2/dt_3}\\
\frac{dt_{\rm 1}}{dt_{\rm 3}}&=&1-\frac{1}{c}\frac{d}{dt_{\rm 3}}\Big[{\cal R}_{\rm AB}({\vec x}_{\rm A1},{\vec x}_{\rm B2})+{\cal R}_{\rm BA}({\vec x}_{\rm B2},{\vec x}_{\rm A3})
\Big]-\frac{1}{c}\frac{d}{dt_{\rm 3}}(\delta^{\rm atm}_{\rm A12}+\delta^{\rm atm}_{\rm A23}),
\label{eq:dt_A/dt_B0}
\end{eqnarray}
where we introduced atmospheric delay $\delta^{\rm atm}_{\rm A12}$ for the signal from station A on the up-link and atmospheric delays $\delta^{\rm atm}_{\rm A23}, \delta^{\rm atm}_{\rm B23}$ for the downlink leg of the two-way signal originating at station $A$ and the downlink signal originating at $B$, respectively. This allows us to account for the fact that atmospheric delay is frequency dependent effect increasing the length of the propagation path in the atmosphere. Note that due to the relativistic frequency shift, atmospheric delay is different for uplinks and downlinks.

The results given by Eqs.~(\ref{eq:dt2/dt_3})--(\ref{eq:dt_A/dt_B0}) allow us to present (\ref{eq:Doppler-canceled}), describing the frequency difference between the signal transmitted from $B$ and received at $A$, comparing it to the  local oscillator at $A$, as
{}
\begin{eqnarray}
\delta\eta^{\rm comp}_f&=&
\frac{1}{2c}
\frac{d}{dt_{\rm 3}}\Big[{\cal R}_{\rm AB}({\vec x}_{\rm A1},{\vec x}_{\rm B2})-{\cal R}_{\rm BA}({\vec x}_{\rm B2},{\vec x}_{\rm A3})\Big]+\frac{1}{c}\frac{d}{dt_{\rm 3}}\Big({\textstyle\frac{1}{2}}(\delta^{\rm atm}_{\rm A12}+\delta^{\rm atm}_{\rm A23})-\delta^{\rm atm}_{\rm B23}\Big)+\nonumber\\
&+&
\Big[\Big(1+\delta\dot{\hat\tau}_{\rm B}+\delta\dot\tau^{\rm r}_{\rm B}-\delta\dot{\hat\tau}_{\rm A}-\delta\dot\tau^{\rm r}_{\rm A}\Big)\Big(\frac{d\tau_{\rm B}}{dt}\Big)_{t_2}\Big(\frac{d\tau_{\rm A}}{dt}\Big)^{-1}_{t_3}-1\Big]\Big(1-\frac{1}{c}\frac{d}{dt_{\rm 3}}\Big[{\cal R}_{\rm BA}({\vec x}_{\rm B2},{\vec x}_{\rm A3})\Big]-\frac{1}{c}\frac{d\delta^{\rm atm}_{\rm B23}}{dt_{\rm 3}}\Big)-
\nonumber\\
&-&
\frac{1}{2}\Big[\Big(1+\delta\dot\tau^{\rm e}_{\rm A}-\delta\dot\tau^{\rm r}_{\rm A}\Big)\Big(\frac{d\tau_{\rm A}}{dt}\Big)_{t_1}\Big(\frac{d\tau_{\rm A}}{dt}\Big)^{-1}_{t_3}-1\Big]\times\nonumber\\
&&\hskip 100pt\times
\Big(1-\frac{1}{c}\frac{d}{dt_{\rm 3}}\Big[{\cal R}_{\rm AB}({\vec x}_{\rm A1},{\vec x}_{\rm B2})+{\cal R}_{\rm BA}({\vec x}_{\rm B2},{\vec x}_{\rm A3})
\Big]-\frac{1}{c}\frac{d}{dt_{\rm 3}}(\delta^{\rm atm}_{\rm A12}+\delta^{\rm atm}_{\rm A23})\Big).
\label{eq:Doppler-can_comp}
\end{eqnarray}

The resulting expression is valid for arbitrary worldlines of  $A$ and $B$. Although (\ref{eq:Doppler-can_comp}) contains all three values of time, $t_{\rm 1}, t_{\rm 2}, t_{\rm 3}$, any two of these values are determined by the third. We choose the time of final reception at $A$, $t_{\rm 3}$, as the independent variable. The values of $t_{\rm 1}$ and $t_{\rm 2}$ may be explicitly expressed via $t_{\rm 3}$ as $t_{\rm 1}(t_3)$ and $t_{\rm 2}(t_3)$ by applying the transformations (\ref{eq:synchron}) and (\ref{eq:transm-time:ABA}). Furthermore, (\ref{eq:proper-coord-t+}) and (\ref{eq:proper-coord-t}) relate the coordinate time $t_3$ to the proper time $\tau_{\rm A3}$ and can be integrated to determine $t_3=t_3(\tau_{\rm A3})$ and vice versa.

To simplify the model (\ref{eq:Doppler-can_comp}), we evaluate the terms on the right-hand side of this equation using the ACES configuration. To do this, we estimate the following ratio:
{}
\begin{eqnarray}
1-\Big(\frac{d\tau_{\rm B}}{dt}\Big)_{t_2}\Big(\frac{d\tau_{\rm A}}{dt}\Big)^{-1}_{t_3}&=&
\frac{1}{c^2}\Big({\textstyle\frac{1}{2}}\big({{\vec v}^2_{\rm B2}-{\vec v}^2_{\rm A3}}\big)+U_{\rm E}({\vec x}_{\rm B2})-U_{\rm E}({\vec x}_{\rm A3})\Big)+{\cal O}(c^{-4})\simeq 3.675\times 10^{-10}.
\label{eq:tau-comb-AB}
\end{eqnarray}
This term is significant and must be kept in the model. The $1/c^4$ terms in (\ref{eq:tau-comb-AB}) produce contributions of the order of $\sim 5\times 10^{-19}$, which is beyond the ACES capability and may be omitted. Next, using expansions presented in the Appendix B of \cite{Turyshev:2014dea} we evaluate the term
{}
\begin{eqnarray}
\frac{1}{c}\frac{d}{dt_{\rm 3}}\Big[{\cal R}_{\rm BA}({\vec x}_{\rm B2},{\vec x}_{\rm A3})\Big]&=&
\frac{1}{c}\frac{d}{dt_{\rm 3}}|{\vec x}_{\rm A}(t_3) - {\vec x}_{\rm B}(t_2)|+{\cal O}(c^{-3})=\frac{1}{c}({\vec n}_{\rm A3B2}\cdot {\vec v}_{\rm A3B2})+{\cal O}(c^{-2})=\nonumber\\
&=&2.71\times10^{-5}+{\cal O}(7\times 10^{-10}),
\label{eq:r_AB}
\end{eqnarray}
where ${\vec r}_{\rm A3B2}={\vec r}_{\rm B2}-{\vec r}_{\rm A3}$, $r_{\rm A3B2}=|{\vec r}_{\rm A3B2}|$, ${\vec n}_{\rm A3B2}={\vec r}_{\rm A3B2}/r_{\rm A3B2}$, and ${\vec v}_{\rm A3B2}={\vec v}_{\rm B2}-{\vec v}_{\rm A3}$
Thus, this term also must be included in the model. However, the atmospheric contribution may be omitted from this term. Indeed, atmospheric delay is $\delta^{\rm atm}_{12}\sim 2$~m and, even if it changes at a rate of $\sim 2$~m/s, the resulting term $c^{-1}\delta^{\rm atm}_{12}=6.67\times 10^{-9}$, multiplied by (\ref{eq:tau-comb-AB}), produces a negligible contribution.

Similarly we evaluate the factor in the square brackets in second term of (\ref{eq:Doppler-can_comp}). Thus, using  (\ref{eq:proper-coord-t}) we have:
\begin{equation}
1-\Big(\frac{d\tau_{\rm A}}{dt}\Big)_{t_1}\Big(\frac{d\tau_{\rm A}}{dt}\Big)_{t_3}^{-1}= \frac{1}{c^2}\frac{d}{dt}\Big[\frac{{\vec v}^2_{\rm A}}{2}+U_{\rm E}({\vec x}_{\rm A})\Big]\Delta t_{13}+{\cal O}(\Delta t^2_{13}, c^{-4}),
\label{eq:tau-comb-A}
\end{equation}
where $\Delta t_{13}=t_{\rm 3}-t_{\rm 1}$. The magnitude of the RHS can be easily evaluated. For the time of transmission in the two-way case, we get $\Delta t_{13}\sim 2d_{\rm AB}/c$, with maximal value of $\Delta t^{\rm max}_{13}=d_{\rm AB}^{\rm max}/c=2c^{-1}\sqrt{(r_{\rm E}+h_{\rm ISS})^2 -r_{\rm E}^2}\approx2c^{-1}\sqrt{2r_{\rm E}h_{\rm ISS}}=15.2$~ms. Therefore, the RHS of (\ref{eq:tau-comb-A}) has the magnitude
{}
\begin{eqnarray}
\frac{d}{dt}\Big[{\frac{{\vec v}^2_{\rm A}}{2}}+\frac{GM}{r_{\rm A}}\Big]\cdot\frac{2d_{\rm AB}}{c^3}&\leq&
1.3\times 10^{-18},
\label{eq:tau_Ad}
\end{eqnarray}
which is negligible for ACES, allowing us to drop from (\ref{eq:Doppler-can_comp}) the combination of terms proportional to (\ref{eq:tau-comb-A}).

To simplify the third term in (\ref{eq:Doppler-can_comp}), we need to express time $t_1$ via $t_3$, which can be done by using (\ref{eq:transm-time:ABA}):
{}
\begin{eqnarray}
t_1=t_3-\frac{2}{c}r_{\rm A3B2}-\frac{2}{c^2}({\vec r}_{\rm A3B2}\cdot {\vec v}_{\rm A3})+{\cal O}(c^{-3}),
\label{eq:t1-t3}
\end{eqnarray}
Next, using formulae from Appendix B of \cite{Turyshev:2014dea}, the third term in (\ref{eq:Doppler-can_comp}) takes the form:
{}
\begin{eqnarray}
{\cal R}_{\rm AB}({\vec x}_{\rm A1},{\vec x}_{\rm B2})-{\cal R}_{\rm BA}({\vec x}_{\rm B2},{\vec x}_{\rm A3})=\frac{2}{c}({\vec r}_{\rm A3B2}\cdot {\vec v}_{\rm A3})+\frac{2r_{\rm A3B2}}{c^2}\Big({\vec v}_{\rm A3}^2-({\vec r}_{\rm A3B2}\cdot{\vec a}_{\rm A3})\Big)+{\cal O}(c^{-3}),
\label{eq:R-R}
\end{eqnarray}
where ${\vec a}_{\rm A}=\dot{\vec v}_{\rm A}$ is the acceleration of station A. To take the derivative $d/dt_3$ from (\ref{eq:R-R}) we first need to account that $t_2$ and a faction of $t_3$. Thus, using (\ref{eq:dt2/dt_3}), with sufficient accuracy, we have
{}
\begin{eqnarray}
\frac{dt_{\rm 2}}{dt_{\rm 3}}=1-\frac{1}{c}({\vec n}_{\rm A3B2}\cdot {\vec v}_{\rm A3B2})+{\cal O}(c^{-2}).
\label{eq:dt2/dt_3*}
\end{eqnarray}
As a result, the third term in (\ref{eq:Doppler-can_comp}) takes the form:
{}
\begin{eqnarray}
\frac{1}{2c}
\frac{d}{dt_{\rm 3}}\Big[{\cal R}_{\rm AB}({\vec x}_{\rm A1},{\vec x}_{\rm B2})&-&
{\cal R}_{\rm BA}({\vec x}_{\rm B2},{\vec x}_{\rm A3})\Big]=\frac{1}{c^2}\Big(({\vec v}_{\rm A3B2}\cdot {\vec v}_{\rm A3})+({\vec r}_{\rm A3B2}\cdot {\vec a}_{\rm A3})\Big)\Big(1-\frac{1}{c}({\vec n}_{\rm A3B2}\cdot {\vec v}_{\rm A3B2})\Big)+\nonumber\\
&+&\frac{1}{c^3}r_{\rm A3B2}\Big(\big((3{\vec v}_{\rm A3}-{\vec v}_{\rm B2})\cdot{\vec a}_{\rm A3}\big)-({\vec r}_{\rm A3B2}\cdot\dot{\vec a}_{\rm A3})\Big)+{\cal O}(c^{-3}).
\label{eq:R-R*}
\end{eqnarray}

As a result, Eq.~(\ref{eq:Doppler-can_comp}) may be written as
{}
\begin{eqnarray}
\delta\eta^{\rm comp}_f&=&-
\frac{1}{c^2}\Big({\textstyle\frac{1}{2}}{{\vec v}^2_{\rm A3B2}}+U_{\rm E}({\vec x}_{\rm B2})-U_{\rm E}({\vec x}_{\rm A3})-({\vec r}_{\rm A3B2}\cdot {\vec a}_{\rm A3})\Big)
\Big(1-\frac{1}{c}({\vec n}_{\rm A3B2}\cdot {\vec v}_{\rm A3B2})\Big)+\nonumber\\
&&\hskip 0pt +~
\frac{1}{c^3}r_{\rm A3B2}\Big(\big((3{\vec v}_{\rm A3}-{\vec v}_{\rm B2})\cdot{\vec a}_{\rm A3}\big)-({\vec r}_{\rm A3B2}\cdot\dot{\vec a}_{\rm A3})\Big)+\frac{1}{c}\dot\rho^{\rm atm}_{\rm 3w}+\nonumber\\
&&\hskip 0pt +~
\Big(\delta\dot{\hat\tau}_{\rm B}+\delta\dot\tau^{\rm r}_{\rm B}-\delta\dot{\hat\tau}_{\rm A}-\delta\dot\tau^{\rm e}_{\rm A}\Big)\Big(1-\frac{1}{c}({\vec n}_{\rm A3B2}\cdot {\vec v}_{\rm A3B2})\Big)+\frac{1}{2}\Big(\delta\dot\tau^{\rm e}_{\rm A}-\delta\dot\tau^{\rm r}_{\rm A}\Big)
+{\cal O}(1.3\times 10^{-18}).~~
\label{eq:Dopp-can_comp**+}
\end{eqnarray}
This quantity is fully determined by the worldlines of the two stations at various moments of signal propagation. We introduced $\rho^{\rm atm}_{\rm 3w}={\textstyle\frac{1}{2}}(\delta^{\rm atm}_{\rm A12}+\delta^{\rm atm}_{\rm A23})-\delta^{\rm atm}_{\rm B23}$, which denotes temporal change in the asymmetry of the atmospheric delay between down- and uplinks during the combination of a two-way and a one-way communication link used to cancel the first-order Doppler shifts. We can see that the first-order Doppler cancelation scheme given by Eqs.~(\ref{eq:Doppler-can_obs=})--(\ref{eq:Doppler-can_comp*}) also greatly reduces contributions associated with various noise sources, especially atmospheric noise.  Originally, with accuracy up to $1/c^2$, Eq.~(\ref{eq:Dopp-can_comp**+}) was developed for the {\em Gravity Probe A} experiment \cite{Vessot1980}, which was later updated for ACES to include the $1/c^2$ terms in \cite{Blanchet-etal:2001}. Our results are similar, but they were developed using a different approach, which allows for a significant generalization.

As a last step, we express all the quantities in (\ref{eq:Dopp-can_comp**+}) using the proper time of final reception, $\tau_{\rm A3}$, corresponding to coordinate time $t_3$. This can be done by using the relation between $t_2$ and $t_3$, which to sufficient order is
{}
\begin{eqnarray}
t_2=t_3-\frac{1}{c}r_{\rm AB}+{\cal O}(c^{-2}).
\label{eq:t2-t3}
\end{eqnarray}
As a result, we are able to express the model (\ref{eq:Dopp-can_comp**+}) for the Doppler-canceled gravitational redshift in its most convenient form:
{}
\begin{eqnarray}
\delta\eta^{\rm comp}_f&=&-
\frac{1}{c^2}\Big({\textstyle\frac{1}{2}}{{\vec v}^2_{\rm AB}}+U_{\rm E}({\vec x}_{\rm B})-U_{\rm E}({\vec x}_{\rm A})-({\vec r}_{\rm AB}\cdot {\vec a}_{\rm A})\Big)
\Big(1-\frac{1}{c}({\vec n}_{\rm AB}\cdot {\vec v}_{\rm AB})\Big)-\nonumber\\
&&\hskip 0pt -
\frac{1}{c^3}r_{\rm AB}\Big(2({\vec v}_{\rm AB}\cdot{\vec a}_{\rm A})+({\vec v}_{\rm A}\cdot{\vec a}_{\rm AB})+({\vec r}_{\rm AB}\cdot\dot{\vec a}_{\rm A})\Big)+\frac{1}{c}\dot\rho^{\rm atm}_{\rm 3w}+\nonumber\\
&&\hskip 0pt +~
\Big(\delta\dot{\hat\tau}_{\rm B}+\delta\dot\tau^{\rm r}_{\rm B}-\delta\dot{\hat\tau}_{\rm A}-\delta\dot\tau^{\rm e}_{\rm A}\Big)\Big(1-\frac{1}{c}({\vec n}_{\rm AB}\cdot {\vec v}_{\rm AB})\Big)+\frac{1}{2}\Big(\delta\dot\tau^{\rm e}_{\rm A}-\delta\dot\tau^{\rm r}_{\rm A}\Big)
+{\cal O}(1.3\times 10^{-18}),
\label{eq:Dopp-can_comp***+}
\end{eqnarray}
where ${\vec a}_{\rm AB}={\vec a}_{\rm B}-{\vec a}_{\rm A}$ and all the quantities are expressed in terms of $t_3$. Ultimately, ACES will be able to test the gravitational redshift to an uncertainty level of $<2.0\times 10^{-6}$ in 10 days of integration time. Therefore, the contribution of each term in (\ref{eq:Dopp-can_comp***+}) must be known to an appropriate accuracy. We address these questions and the related precision of the ISS orbit in the following subsection. 

Note that one can also use timing observables to investigate the relative frequency stability between the clocks on the ground and those is space. In Sec.~\ref{sec:timing_observe}}, we introduced  desynchronization between the two clocks with observable, $\delta\tau_{\rm AB}(t)$, given by (\ref{eq:de-offest-obs_ACES}) and the model (\ref{eq:de-offest-comp2_ACES}). The local time stability of these measurements may allow for a precise estimation of a time drift over a short duration (around 1 min), according to this approximation of the derivative:
{}
\begin{eqnarray}
\frac{\delta f^{\rm obs}_{\rm AB}}{f_0}&=&\frac{1}{T}\Big\{\delta\tau^{\rm obs}_{\rm AB}(t+{\textstyle\frac{1}{2}}T)-\delta\tau^{\rm obs}_{\rm AB}(t-{\textstyle\frac{1}{2}}T)\Big\}\quad~ {\rm and}\quad~
\frac{\delta f^{\rm comp}_{\rm AB}}{f_0}=\frac{1}{T}\Big\{\delta\tau^{\rm comp}_{\rm AB}(t+{\textstyle\frac{1}{2}}T)-\delta\tau^{\rm comp}_{\rm AB}(t-{\textstyle\frac{1}{2}}T)\Big\}.
\label{eq:de-offest-obs_ACES-freq}
\end{eqnarray}
These quantities  correspond to a relative frequency offset, $\delta f^{\rm }_{\rm AB}$, seen by the station A during the time interval $T$ with respect to the nominal frequency $f_0$ of the oscillator at $A$. Such a quantity may allow the use of laser ranging data from the ELT experiment for an independent verification of $\delta f^{\rm }_{\rm AB}/f_0$ established by the MWL.

\subsection{Precision of ISS navigation}

Now we can address the question of navigational precision for the ISS, which is needed to satisfy the ACES requirements based on the anticipated precision of the red-shift experiment. To do that, we consider the proper-to-coordinate time transformation for the clock on the ISS that is given by (\ref{eq:prop-coord-time-J2*=}). We begin with presenting this equation in terms of the Keplerian elements of the orbit of the ISS including the semi-major axis, $a$, eccentric anomaly, ${\cal E}={\cal M}+e\sin{\cal E}$ (with ${\cal M}$ being the mean anomaly), eccentricity, $e$, orbital radius, $r=a(1-e\cos{\cal E})$.

Collecting all the terms relevant to establishing the orbit of the ACES clocks, we have:
{}
\begin{eqnarray}
\frac{d\tau_{\rm B}}{dt}-1&=&-
\frac{GM_{\rm E}}{c^2a_0}\Big\{{\textstyle\frac{3}{2}}+{\textstyle\frac{7}{2}}J_{\rm 2E}\Big[\frac{R_{\rm E}}{a_0}\Big]^2\big(1-{\textstyle\frac{3}{2}}\sin^2 i_0\big)+2\Big(1-{\textstyle\frac{3}{2}}J_{\rm 2E}\Big[\frac{R_{\rm E}}{a_0}\Big]^2\big(1-{\textstyle\frac{3}{2}}\sin^2 i_0\big)\Big)e_0\cos{\cal E}_0+\nonumber\\
&&
\hskip -0pt+~
2e^2_0\cos^2{\cal E}_0+
J_{\rm 2E}\Big[\frac{R_{\rm E}}{a_0}\Big]^2\sin^2 i_0\cos2(\omega_0+u)-
\cos\theta_{\rm A_0}\frac{y_{\rm A0}}{a_0}+\pi \frac{c_d A}{m}\rho_{\rm atm}a_0\Big\}+
{\cal O}(4.4\times10^{-17}),~~~
\label{eq:perturb-all}
\end{eqnarray}
where $i_0$ is the inclination, $\omega$ is the altitude of perigee and $u$ is the true anomaly. Also, the subscript $0$ refers to an unperturbed quantity and the error term is set by the lunar tides.

The first term in Eq.~(\ref{eq:perturb-all}) evaluates to $9.83\times10^{-10}$ and is responsible for the largest frequency shift for a clock on Keplerian orbit around the Earth. The second term, which is of the order of $1.73\times10^{-13}$, represents perturbation due the Earth's oblateness. The third term is the eccentricity correction, which was evaluated to be $7.86\times10^{-13}$. Note that oblateness contributes a $\sim1\times 10^{-16}$ correction to this term. The fourth term is a quadratic eccentricity correction with a magnitude of $4.72\times10^{-16}$. The fifth term has an amplitude $3.86\times10^{-13}$. The sixth term amounts to $\sim 3.39\times 10^{-16}$ and is due to the shift of the ACES position with respect to the center of origin of the SCRS on the ISS. The last term has the magnitude of up to $\sim 1.15\times 10^{-14}$ is due to atmospheric drag. The contributions of all these terms are significant at the expected level of accuracy of ACES.

Considering the first term in Eq.~(\ref{eq:perturb-all}) (or, similarly, the first two terms in (\ref{eq:Dopp-can_comp**+}) for the red-shift experiment), we see that during a 90-minute ISS orbit, the altitude of the space station must be known to
{}
\begin{eqnarray}
\delta a_0=a_0\Big[\frac{3GM_{\rm E}}{2c^2a_0}\Big]^{-1}\delta\Big(\frac{d\tau_{\rm A}}{dt}\Big)\leq6.89\times 10^{15}\, \delta\Big(\frac{d\tau_{\rm A}}{dt}\Big)~{\rm m}.
\label{eq:req_pos}
\end{eqnarray}
It is anticipated that during an integration time of $10^3$~s, the frequency stability of the ACES clock is dominated by the stability of the MWL and should be better than $\delta f/f=\delta(d\tau_{\rm A}/dt)=2.1\times10^{-15}$ (Table~\ref{tb:caps}). To estimate the size of allowable navigational error, we rely on the ``rule of thumb'' stating that no individual contribution to the total error shall exceed 10\% of the total error. This implies that the position of the ISS on its orbit around the Earth should be known with accuracy better than  $\delta a_0\lesssim\sqrt{0.1}\cdot14.47~{\rm m}\simeq 4.58~{\rm m}$ (i.e., similar to \cite{Duchayne-etal:2009, Ashby-etal-2014}). As the orbital plane of the ISS precesses with each revolution, it is necessary to track the orbital changes on a permanent basis. Therefore, a combination of several methods of orbit determination may be needed, including the use of traditional radio-tracking in conjunction with GPS and SLR (for instance, as part of the ELT experiment).

The presence of the $J_2$ term in (\ref{eq:perturb-all}) may be accounted for within a perturbation theory. Perturbations of the ISS orbits due to Earth's quadrupole are a significant fraction of the change in semi-major axis associated with the corresponding orbit change. One needs to estimate the effect of the Earth's quadrupole moment on the orbital elements of a Keplerian orbit of the space station and also on the change in frequency induced by an orbit change. Accounting for the perturbation in Keplerian orbital elements of the ISS orbit including the semi-major axis, $a$, eccentric anomaly, ${\cal E}={\cal M}+e\sin{\cal E}$ (with ${\cal M}$ being the mean anomaly), eccentricity, $e$, orbital radius, $r=a(1-e\cos{\cal E})$, we can compute perturbations to each of the terms $v_{\rm A}^2$,  in ${GM_{\rm E}}/{r_{\rm A}}$ and the quadrupole term in (\ref{eq:prop-coord-time-J2*=}) (or, equivalently, in Eq.~(\ref{eq:perturb-all})). The corresponding calculations are lengthy but straightforward, and are well-known. Here we present only the final relevant result to the frequency shift of the ACES clock induced by the orbital parameters of the ISS. Considering only the last periodic term in (\ref{eq:perturb-all}), the additional time elapsed for an orbiting clock may be given as
{}
\begin{eqnarray}
\Delta t_{J_2}=-\frac{GM_{\rm E}}{c^2a_0}J_{\rm 2E}\Big[\frac{R_{\rm E}}{a_0}\Big]^2\sin^2 i_0\int_{\rm path}dt
\cos2(\omega_0+nt),
\label{eq:perturbJ2}
\end{eqnarray}
where the true anomaly was replaced by $u=nt$, with $n=(GM_{\rm E}/a^3_0)^{\frac{1}{2}}$ being the approximate mean motion of the ISS. Integrating and dropping the constant of integration (assuming as usual that such constant time offsets are lumped with other contributions) gives the periodic relativistic effect on the elapsed time of a clock due to Earth's quadrupole moment:
{}
\begin{eqnarray}
\Delta t_{J_2}=-\frac{\sqrt{GM_{\rm E}a_0}}{2c^2}
J_{\rm 2E}\Big[\frac{R_{\rm E}}{a_0}\Big]^2\sin^2 i_0\sin2(\omega_0+nt).
\label{eq:perturbJ2a}
\end{eqnarray}
The phase of this effect is zero when the ISS passes through Earth's equatorial plane going northwards. The magnitude of this effect $\Delta t_{J_2\rm ACES} = 1.70\times 10^{-10}$~s, which is significant for the ACES experiment. Contributions of higher zonal harmonics in the Earth's gravitational potential are beyond the ACES sensitivity and  may be neglected.

Next, we consider the frequency shift due to the ACES position on the ISS. By parameterizing the ACES position vector with respect to the SCRS as ${\vec y}_{\rm A0}=y_{\rm A0}{\vec n}_{\rm A0}$, we may present the relevant term in (\ref{eq:prop-coord-time-J2*=}) in the form
{}
\begin{eqnarray}
\Big(\frac{d\tau_{\rm B}}{dt}-1\Big)_{\rm loc}&=&-
\frac{1}{c^2}({\vec a}_{\rm ISS_0}\cdot{\vec y}_{\rm A0})
= \frac{GM_{\rm E}}{c^2a_0^2}y_{\rm A0}({\vec n}_{\rm ISS_0}\cdot{\vec n}_{\rm A0})\leq2.91\times 10^{-15}\cos\theta_{\rm A_0},
\label{eq:perturb_ad-a0}
\end{eqnarray}
where  $\cos\theta_{\rm A_0}=({\vec n}_{\rm ISS_0}\cdot{\vec n}_{\rm A0})$ with ${\vec n}_{\rm ISS}$ being the unit vector in the direction to the ISS from GCRS.

As the reference systems introduced on the ISS accelerate in the Earth's gravity field, their clocks follow different worldlines separated by a small, but finite distance. Such a separation leads to an acceleration-induced redshift between the clocks at the origins of the SCRS and the ACRS. Instrumentally, this effect represents a constant bias in the ACES clocks compared to the time in the SCRS. Because of the small eccentricity of the ISS orbit of $e=0.006$ (see Table \ref{tb:params}), any variability in (\ref{eq:perturb_ad-a0}) will be at least $e$-times smaller and, thus, insignificant. The ACES package will be located on the exterior surface of the Columbus module being at $\sim 30$~m from the ISS center of gravity. In addition, the package will be at $\sim3.5$~m in the nadir direction from that point. This position results in $\cos\theta_{\rm A_0}\simeq 0.12$, thereby reducing the effect (\ref{eq:perturb_ad-a0}) to $3.39\times 10^{-16}$, which is small, but still significant for ACES.

The near-Earth environment also contributes to the clock rate, especially the atmospheric drag which depends on the air density, $\rho_{\rm atm}$, at the ISS altitude. Thus, the density of the upper Earth's atmosphere must be monitored. From Eq.~(\ref{eq:perturb-all}) suggests that one needs to know this quantity
{}
\begin{eqnarray}
\frac{\delta \rho_{\rm atm}}{\rho_{\rm atm}}=\Big[\frac{GM_{\rm E}}{c^2}\pi \frac{c_d A}{m}\rho_{\rm atm}\Big]^{-1}\delta\Big(\frac{d\tau_{\rm B}}{dt}\Big)=8.80\times10^{13}\,\delta\Big(\frac{d\tau_{\rm B}}{dt}\Big),
\label{eq:req_ad}
\end{eqnarray}
which implies that the atmospheric conditions along the orbital track of the ISS must be known to $\sim 18\%$, which may be challenging as this effect is dissipative and changes each orbital pass. Furthermore, due to unavoidable orbital boosts needed to raise the altitude of the ISS, the air density will also change.

Atmospheric drag results in the lowering of the semi-major axis of the ISS, leading to a change of the gravitational potential and velocity of the station. This change may indirectly affect relativistic observables. Assuming a circular orbit for the ISS, we approximate the changes in the semi-major axis, $a$, and orbital velocity, ${\vec v}$, using the equations
\begin{equation}
[\Delta a_{\rm ad}]_{\tt rev}=-2\pi\frac{c_dA}{m}\rho_{\rm atm}a^2_0,
\hskip 30pt
[\Delta v_{\rm ad}]_{\tt rev}=\pi \frac{c_d A}{m}\rho_{\rm atm} a_0v_0,
 \label{eq:ad-av}
\end{equation}
where $a_0$ is the initial length of the semi-major axis and $v_0$ is the initial velocity. For the ISS, we determine that during each revolution, atmospheric drag causes the station's altitude and velocity to change by up to $-18.4$~m and $0.01$~m/s, respectively, during the period with mean solar activity; and by up to $-239$~m and $0.13$~m/s during extreme solar activity. These numbers are of course worst-case estimates, calculated by assuming that the entire area of the solar panels contributes to the effect (i.e., that the solar panels are perpendicular to the velocity vector with respect to the atmosphere). Nevertheless, to compensate for this drop in orbital velocity, the ISS periodically has to re-boost and regain velocity and orbital altitude.

Orbital changes due to atmospheric drag (\ref{eq:ad-av}) will have a direct impact on the relativistic time and frequency observables of ACES. Taking, for instance, the velocity change per revolution $\Delta v_{\tt rev}$ (given by the second equation in (\ref{eq:ad-av})), from (\ref{eq:prop-coord-time-J2*=}) we see that, depending on the solar activity, these changes produce the contribution to proper-to-coordinate time of the order of $(d\tau_{\rm A}/dt)_{\rm ad}$ and, as a result, its effect on frequency stability may be given as:
{}
\begin{eqnarray}
\Big(1-\frac{d\tau_{\rm B}}{dt}\Big)_{\rm ad}&=&
\frac{1}{c^2}({\vec v}_{\rm ISS_0}\cdot{\vec v}_{\rm ad})_{\rm rev}
\leq\pi \frac{c_d A}{mc^2}\rho_{\rm atm} a_0v^2_0=
\frac{GM_{\rm E}}{c^2}\pi \frac{c_d A}{m}\rho_{\rm atm}\simeq(0.89-11.54)\times 10^{-15}.~~~~~
\label{eq:perturb_ad-vel}
\end{eqnarray}

Although there is no explicit dependence of this effect on the ISS orbit, there is implicit dependence as the air density at the ISS orbit is a function of its altitude and orbital inclination. Atmospheric drag results in the loss of the ISS orbital altitude, causing changes of the gravitational potential at the clock's location. Such an altitude decrease results in the change of the clock rate per orbital revolution at the level of $\big(1-{d\tau_{\rm B}}/{dt}\big)_{\rm ad}=c^{-2}\nabla U_{\rm E}[\Delta a_{\rm ad}]_{\rm rev}\simeq (1.78-23.11)\times10^{-15}$.

An order-of-magnitude improvement in the contribution of position errors to the ACES' frequency transfer stability to $\delta f/f=1\times 10^{-16}$ would put even tougher requirements on the knowledge of the ISS position and the air density at its altitude. In fact,  in accord with (\ref{eq:req_pos}) and (\ref{eq:req_ad}), one would have to require that the geocentric position of the ISS and the air density for each orbit must be known to 0.2~m and 2\%, correspondingly. If the orbital parameters of the ISS would stay constant over many days, these new requirements would not matter. In that case, the corresponding effects would contributions only constant once-per-orbit terms, which could be estimated and removed in the data analysis.  However, we know that this is not true \cite{Montenbruck-etal:2011} and the trajectory of the space station changes significantly each orbit. Therefore, finding ways to mitigate the related uncertainty deserves further study.

\section{Conclusions and recommendations}
\label{sec:conc}

We considered the formulation of a relativistic model for the observables of the ACES mission. We derived an analytic expression that characterizes the process of forming the frequency comparison observables. This material can be used to improve the accuracy of modeling of the ACES fundamental observables.

We presented a hierarchy of relativistic coordinate reference frames that are needed to ACES. We introduced the geocentric (GCRS), topocentric (TCRS), spacecraft (SCRS), and ACES-centric coordinate reference systems, together with the structure of the corresponding metric tensors in each of these systems and the form of the proper relativistic gravitational potentials---all presented at the accuracy required for ACES. We presented the rules for transforming time and position measurements between the reference frames involved. We demonstrated, by meticulously computing all possible forces, that contrary to initial expectations the ACES reference frame is pseudo-inertial at the level of accuracy required by the ACES experiment. We also emphasized the need to recognize the importance of the barycentric (BCRS) reference frame for proper modeling of the solar gravity potential at the ACES' location.

We considered the model for the relativistic observables of the ACES experiment. The currently implemented proper-to-coordinate time transformation is a concern as it accounts only for the monopole contribution of the Earth's gravity field omitting higher multipoles. As such, the current model is accurate only up to $3.86\times 10^{-13}$ for ACES. Including the Earth's oblateness $J_2$ improved the relation between proper and coordinate times. In the improved model, $J_2$ is responsible for a periodic effect of the order of 170 ps (or $\sim 10.2$~cm peak-to-peak) for ACES.

After accounting for the Earth's oblateness $J_2$, the error term in the updated model for ACES is at the level of $3\times 10^{-16}$, the limit set by the higher other gravity harmonics in the gravitational potential of the extended Earth. The conventional form of the light time solution is accurate up to the order of $\sim 2\times 10^{-15}$~s (due to the Earth's $J_2$ coefficient) and, thus, it is adequate for ACES. The usual form for relativistic terms in the spacecraft equations of motion are also adequate for the task. They are accurate up to the contribution coming from relativistic term due to Earth's oblateness \cite{Will-2014}, which was evaluated to be of the order of $\sim 2.85\times 10^{-11}~{\rm m/s}^2$ for a LEO spacecraft, which may be important for some high-precision orbit determination in the near future. However, the presence of the nongravitational forces, especially atmospheric drag, limits the equation to only $\sim 1.0\times 10^{-7}~{\rm m/s}^2$ (from (\ref{eq:sr-acc})) resulting in (\ref{eq:eqm-ISS-real}), which is adequate for the anticipated positional accuracy needed for ACES.

In a practical sense, the small relativistic terms that we calculated are easily absorbed into constant and periodic ad-hoc biases that are introduced during data analysis, with no impact whatsoever on mission objectives or the quality of the mission's results. Yet the existence of these terms, and the fact that they are observable at the level of sensitivity of the ACES experiment demonstrate that ACES is already a practical instrument for relativistic geodesy. For future spacecraft that operate at even greater accuracy, accounting for these relativistic terms will be essential.

As we discussed, the ISS is a subject to many nongravitational forces and related torques; therefore, the ISS may not be treated as a free-falling platform. Although any direct effect of nongravitational accelerations on clock rates is negligible, these forces still affect the ACES experimental precision. The effect is indirect and comes from the uncertainties introduced in the estimates of the ISS trajectory. In fact, the ACES experiment must rely on the navigational precision of the ISS to reach its science objectives.

In particular, it is necessary to consider the effect due to attitude variations. While this effect is small, it is larger than that of the atmospheric drag and, thus, it is of concern. If such a contribution to the frequency comparison is of a systematic origin, it is possible to develop a model to calibrate ACES observables and remove such an unwanted effect. However, attitude variations at this level may occur during various mission events (e.g., docking/un-docking of the crew and cargo supply vehicles, configuration changes in the ISS, crew exercise, thruster firings, etc.) It may be necessary to monitor the dynamical environment on the ISS to keep track of those changes in order to be able to account for their likely effects on the ACES clock.

Comparing the newly developed models for time transfer and
gravitational redshift given by (\ref{eq:de-offest-comp2_ACES}) and (\ref{eq:Dopp-can_comp***+}), we see that frequency transfer puts more demanding requirements on the orbit of the space clock. Similar questions were addressed previously under different set of assumptions \cite{Duchayne-etal:2009,Schreiber-etal:2009,Montenbruck-etal:2011,Wermuth-etal:2012}. Our intent here was to explicitly describe the sources of systematic error that could affect the time and frequency transfer measurements on ACES. We developed a set of relativistic models for the ACES observables and were able to dissect the total error into various dynamical contributions. In this work, we found that to satisfy the anticipated stability of the frequency transfer, the geocentric position of the ACES clock and the air density at the ISS altitude must be known to $\sim$4.6~m and 18\%, correspondingly, consistent with earlier results \cite{Duchayne-etal:2009}. We investigate the ways to satisfy these requirements including the use of the advanced facility for high-precision satellite laser ranging being currently developed at JPL and implementation of the new data analysis strategy that relies on processing the clock and navigational data together.

To evaluate navigational needs of ACES, we used an approach typical for deep space navigation, where the clock (or USO) is treated as part of navigational subsystem that has a direct impact on navigation and, thus, on science.  The orbit of a spacecraft in this case is typically determined in a joint data analysis relying on the clock and relevant navigational data – the process that could also be used to estimate various science data parameters. Although, our results are similar to those obtained earlier (i.e., \cite{Duchayne-etal:2009}), our approach is somewhat different from the sequential data analysis flow adopted for ACES, where the GPS navigational solutions are used to calibrate the MWL products, which then are used to process the ACES clock data for science investigations. The formulation presented here allows one to process the ACES clock measurements together with the data from other spacecraft instruments and subsystems. Such a concurrent data analysis capability is new and could be used as a general tool for future ACES-like experiments with clocks of higher precision (for instance, \cite{SOC}).

Concluding, we mention that we begun to address the issues above by developing a comprehensive modeling, simulation, and data analysis software system for ACES. To that extent, we already initiated the development of a simulation system relying in part on existing software suite developed for other missions (i.e., OPALS\footnote{Optical PAyload for Lasercomm Science (OPALS), for details see {\tt http://phaeton.jpl.nasa.gov/external/projects/optical.cfm}}).
We pay special attention to the issues related to ISS navigation and combination of various data-types needed to achieve not only high-precision navigation, but also highly accurate science models. The high-accuracy models for the time transfer and the gravitational redshift experiment developed here will be used
to process the ACES measurements for science data analysis. The corresponding software suite relying on the models obtained here is currently being developed. Preliminary results are encouraging and will be published in subsequent publications.

\begin{acknowledgments}
This work was performed at the Jet Propulsion Laboratory, California Institute of Technology, under a contract with the National Aeronautics and Space Administration.
\end{acknowledgments}

\end{document}